\documentclass[reprint,superscriptaddress,amsmath,amssymb,aps,prb,floatfix,eqsecnum]{revtex4-2}
\usepackage{graphicx}
\usepackage{dcolumn}
\usepackage{mathrsfs}
\usepackage{bm}
\usepackage{subcaption}
\usepackage{amsfonts}
\usepackage{amssymb}
\usepackage{lipsum}
\usepackage[normalem]{ulem}
\usepackage{bbm}
\usepackage{float}
\usepackage{mdframed,xcolor}
\usepackage{xcolor}
\usepackage{caption}
\usepackage{changes}
\usepackage{caption}
\usepackage[normalem]{ulem}
\usepackage{tcolorbox}
\definecolor{myshade}{RGB}{255,235,235}
\usepackage{colortbl}
\usepackage{tikz} 
\usepackage[colorlinks=true,linkcolor=blue, citecolor=red, urlcolor=blue]{hyperref}
\usepackage{url}

\newcommand{\ket}[1]{|#1\rangle}
\newcommand{\bra}[1]{\langle#1|}
\newcommand{\braket}[2]{\langle#1|#2\rangle}
\newcommand{\mel}[3]{\langle#1|#2|#3\rangle}

\newcommand{\Repart}{\operatorname{Re}}
\newcommand{\Impart}{\operatorname{Im}}

\begin{document}
\title{A Gauge-Covariant Geometric Framework for Non-Hermitian Quantum Systems}

\author{Gargi Das}
\email{96gargi@gmail.com, \\gargi.das@bhu.ac.in}
\affiliation{Department of Physics, Banaras Hindu University, Varanasi 221005, INDIA}
\author{Sudhanshu Shekhar$^{1,}$}
\email{shekhar1997sudhanshu@gmail.com, \\sudhanshushekhar@bhu.ac.in
}
\author{Bhabani Prasad Mandal$^{1,}$} 
\email{bhabani.mandal@gmail.com, bhabani@bhu.ac.in}
\noaffiliation
\date{\today}
\begin{abstract}
We develop a comprehensive, gauge-covariant geometric framework for non-Hermitian quantum systems in the quasi-Hermitian regime, that is, the region of parameter space where the non-Hermitian Hamiltonian admits a real spectrum and a positive-definite metric operator. We build this framework by elevating the Dyson map to a central geometric object. This map is the transformation that converts a non-Hermitian Hamiltonian into an equivalent Hermitian one. From it we construct the Dyson connection and decompose it into Hermitian and anti-Hermitian parts, identified respectively as {\it stretching } and  {\it rotation } components. This decomposition cleanly separates the genuine physical metric deformations from the unitary gauge redundancies. Working with manifestly gauge-covariant states, we then derive the complex non-Hermitian Berry phase and the quantum geometric tensor (QGT), and show that the non-Hermitian geometric curvature originates from the non-commutativity of the stretching components at the operator level. We further analyse the geometric singularities near an exceptional point (EP) and uncover a distinct hierarchy of divergences. For a general two-level non-Hermitian model, the quantum metric tensor (QMT) exhibits a leading-order divergence $\sim |\epsilon_\mu|^{-2}$, while the Berry curvature shows a weaker, subleading divergence $\sim |\epsilon_\mu|^{-3/2}$, with $\epsilon_\mu$ denoting the parameter displacement from the EP along an individual parameter axis $\mu$. Finally, we examine physical realizations of this model, including the non-Hermitian Su--Schrieffer--Heeger (SSH) and Hatano--Nelson (HN) models, where exact analytical results confirm the predicted critical scaling laws and illustrate the metric-deformation-driven non-Hermitian geometries.
\end{abstract}

\maketitle

\section{Introduction}
\label{sec:introduction}
%
The geometric formulation of quantum mechanics has profoundly deepened our understanding of physical phenomena. Since the discovery of the Berry phase~\cite{berry1984quantal}, it has become clear that the parameter space of a quantum system carries a rich geometric and topological structure~\cite{simon1983holonomy, shapere1989geometric}, encoded in the quantum geometric tensor (QGT)~\cite{cheng2010quantum, provost1980riemannian, ma2010abelian}. Its imaginary part yields the Berry curvature, which underlies topological invariants and anomalous transport phenomena~\cite{xiao2010berry, hasan2010colloquium, qi2011topological, bansil2016colloquium, nagaosa2010anomalous}, while its real part defines the quantum metric tensor (QMT), which measures the distance between neighbouring quantum states~\cite{cheng2010quantum} and governs quantum phase transitions~\cite{zanardi2007information}, orbital magnetic susceptibility~\cite{gao2015geometrical, raoux2015orbital, piechon2016geometric}, and quantum information geometry~\cite{kolodrubetz2017geometry, campos2007quantum, amari2000methods}. In standard Hermitian quantum mechanics this geometry rests on the rigid, usual Dirac inner product of Hilbert space.\\
This foundation is disrupted in non-Hermitian quantum systems, which have emerged as a major frontier of modern physics~\cite{bender2007making, ashida2020non, moiseyev2011non, el2018non, hajong2026emergence}. Non-Hermiticity arises naturally in open systems with gain and loss~\cite{el2018non, jiang2024tunable, li2022gain, liu2020gain, das2024thermal}, photonic circuits~\cite{liu2020gain, ozdemir2019parity, de2022non, li2023exceptional, parto2020non, xiao2025non, schomerus2013topologically}, classical metamaterials~\cite{ghatak2020observation, li2024experimental, wang2023non, wu2026observation}, and systems with finite quasiparticle lifetimes~\cite{kozii2024non, turkeshi2022entanglement, nagai2020dmft}. A central discovery in this field is that non-Hermitian Hamiltonians can possess entirely real spectra when they respect parity-time ($\mathcal{PT}$) symmetry~\cite{bender1998real, bender1999PT, bender2002complex}. This insight has driven the construction and analysis of a broad class of $\mathcal{PT}$-symmetric models, ranging from quasi-exactly solvable potentials~\cite{khare2009new} and Calogero-type many-body systems~\cite{mandal2012spectral, basu2001exactly} to $\mathcal{PT}$-symmetric field-theoretic extensions~\cite{dwivedi2021higher}, while bringing into focus structural features such as parity violation and reciprocity~\cite{ghatak2015reciprocity}. The reality of the spectrum was later understood through the broader framework of pseudo-Hermiticity~\cite{mostafazadeh2002pseudo, mostafazadeh2002pseudo2, mostafazadeh2002pseudo3} and the earlier notion of quasi-Hermiticity~\cite{scholtz1992quasi}: a quasi-Hermitian Hamiltonian has a real spectrum, a complete set of eigenstates, and admits a positive-definite metric operator relating its distinct left and right eigenvectors~\cite{kretschmer2004quasi}. This positive-definite metric is what makes a consistent quantum-mechanical interpretation possible, but it also forces the use of biorthogonal quantum mechanics~\cite{brody2014biorthogonal}, which deforms the inner product and, as a consequence, demands a careful reformulation of even elementary quantum-mechanical relations such as the uncertainty principle~\cite{shukla2023uncertainty} and the Hellmann--Feynman theorem~\cite{hajong2024hellmann}.\\
This deformation of the inner product is not merely a technical inconvenience: it endows the state space with a non-trivial geometric character in which non-Hermiticity itself can act as a source of curvature. The boundary of the quasi-Hermitian region is marked by exceptional points (EPs)~\cite{heiss2012physics, berry2004physics, leykam2017edge}, where eigenvalues and eigenstates simultaneously coalesce, a defective degeneracy with no Hermitian counterpart~\cite{heiss2012physics, bergholtz2021exceptional}. EPs underlie a wide range of phenomena, including unidirectional invisibility~\cite{lin2011unidirectional}, topological mode switching~\cite{doppler2016dynamically}, bulk Fermi arcs~\cite{zhou2018observation}, enhanced sensitivity in quantum sensors~\cite{wiersig2014enhancing, hodaei2017enhanced, chen2017exceptional}, and unconventional scattering features~\cite{hasan2018new, hasan2020new, ghatak2013various}, and they leave distinct imprints on quantum-information measures such as the entanglement entropy of $\mathcal{PT}$-symmetric models~\cite{modak2021eigenstate, das2026quantum}. Near an EP, the collapse of the eigenspace drives a divergence of geometric quantities, making the region around the EP a natural arena in which non-Hermitian geometry becomes most pronounced~\cite{kartik2025scaling, liao2021experimental, solnyshkov2021quantum}.\\
Extending the geometric framework to this setting has therefore been the subject of intense effort~\cite{garrison1988complex, lieu2018topological, borgnia2020non, yin2018geometrical, chen2024quantum, hu2025quantum}. In the biorthogonal framework, the Berry connection becomes complex~\cite{garrison1988complex, dattoli1990geometrical, longhi2023complex}, and the non-Hermitian QGT acquires complex components and indefinite signatures~\cite{chen2024quantum, zhang2019quantum, cuerda2024observation, fan2020complex}, both predicted and measured near EPs~\cite{liao2021experimental, cuerda2024observation}. A recurring theme in this body of work is that non-Hermiticity behaves geometrically like curvature. This perspective was made especially vivid by the demonstration that non-Hermitian models in flat space can be dual to Hermitian models in curved spaces~\cite{lv2022curving}, and by the vielbein (Hermitization) formalism, which maps the non-trivial Hilbert-space metric of a non-Hermitian system to a Hermitian frame and analyses the resulting parallel transport and curvature~\cite{ju2022einstein, ju2024emergent}. These approaches firmly establish that the metric structure of the Hilbert space is the geometric origin of non-Hermitian phenomena.\\
A persistent difficulty, however, runs through these constructions: the biorthogonal basis is not unique and possesses an intrinsic $GL(N,\mathbb{C})$ gauge freedom, so it is often unclear which geometric features are physical and which are artifacts of the chosen frame. This ambiguity has recently been noted, with the geometry reformulated around the Hilbert-space metric to separate the intrinsic eigenstate geometry from the contribution of the parameter-dependent metric~\cite{arkhipov2026resolving}.\\
We provide a framework in which this metric contribution is not only identified but organized into a well-defined, gauge-covariant geometric object with direct physical meaning, by placing the Dyson map~\cite{ghosh2026twin, dyson1956general, scholtz1992quasi, mostafazadeh2010pseudo} at its centre. The Dyson map is the similarity transformation that converts a non-Hermitian Hamiltonian into an equivalent Hermitian one~\cite{mostafazadeh2010pseudo}, flattening the deformed biorthogonal Hilbert space into a standard orthonormal frame. While it preserves the spectrum, it is not unique: it carries an intrinsic unitary gauge freedom~\cite{mostafazadeh2010pseudo, assis2009non, mostafazadeh2002pseudo}, with different choices embedding the non-Hermitian system into the Hermitian frame in different ways. The geometric role of this parameter-dependent map, and how its internal unitary freedom interacts with the physical metric deformation, has not previously been formulated into a systematic gauge-covariant theory.\\
In this work, we treat the Dyson map as a central geometric object over parameter space rather than a mere similarity transformation. From its parameter dependence we construct the Dyson connection and decompose it canonically into a Hermitian component (stretching) and an anti-Hermitian component (rotation). The stretching component is fixed by the variation of the metric operator and therefore encodes the genuine, metric deformation of the state space, while the rotation component absorbs the unitary redundancy in the choice of Dyson map. This separation is the key step: the decomposition retains the metric contribution as a physical part of the geometry. The stretching sector is gauge-invariant and directly measurable through the resulting amplification or attenuation of state norms. This clean split lets us construct manifestly gauge-covariant states, the dressed derivative state and the stretching fluctuation state, from which we systematically derive the complex non-Hermitian Berry connection and Berry phase, and the non-Hermitian QGT with its indefinite QMT and complex Berry curvature. Using the flatness of the Dyson connection, we further show that the non-Hermitian Berry curvature originates from the non-commutativity of the stretching components, giving a clear algebraic source for non-Hermitian geometric effects. Building on this structure, we analyse the geometric singularities near EPs and uncover a distinct hierarchy of divergences: for a general two-level non-Hermitian model, the QMT diverges as $|\epsilon_\mu|^{-2}$ while the Berry curvature diverges only as $|\epsilon_\mu|^{-3/2}$, where $\epsilon_\mu$ is the parameter displacement from the EP along an individual parameter axis. This general model contains the non-Hermitian Su--Schrieffer--Heeger (SSH)~\cite{yao2018edge} and Hatano--Nelson (HN)~\cite{hatano1996localization} models as exact analytical limits, where all geometric quantities are obtained in closed form and the predicted scaling laws and metric-deformation-driven geometries are explicitly validated.\\
The paper is organized as follows. In Sec.~\ref{sec:geometric_framework}, we introduce the biorthogonal setup and the Dyson map. In Sec.~\ref{sec:decomposition}, we define the Dyson connection and decompose it into its stretching and rotation sectors, constructing the covariant states. In Sec.~\ref{sec:complex_berry_phase} and Sec.~\ref{sec:qgt}, we derive the complex Berry phase and the non-Hermitian QGT, linking their components to the underlying operator field strengths. In Sec.~\ref{sec:ep_singularity}, we analyze the scaling laws and divergence hierarchy near an EP. In Sec.~\ref{sec:example}, we apply the framework to a general two-level non-Hermitian model, which contains the non-Hermitian SSH and HN models as special cases. Finally, we conclude in Sec.~\ref{sec:conclusion}. Supporting derivations are collected in the Appendices: the complete classification of the gauge structure (Appendix~\ref{app:gauge_structure}), the derivation of the gauge-invariant curvature identities (Appendix~\ref{app:identities}), the spectral representation of the off-diagonal matrix elements and covariant states (Appendix~\ref{app:Smu_offdiag}), and a comparison of the right--right, left--left, and left--right forms of the QGT (Appendix~\ref{app:RRLL}).
%
\section{Geometric Framework}
\label{sec:geometric_framework}
\begin{figure}[htpb]
    \centering
    \includegraphics[width=0.95\linewidth]{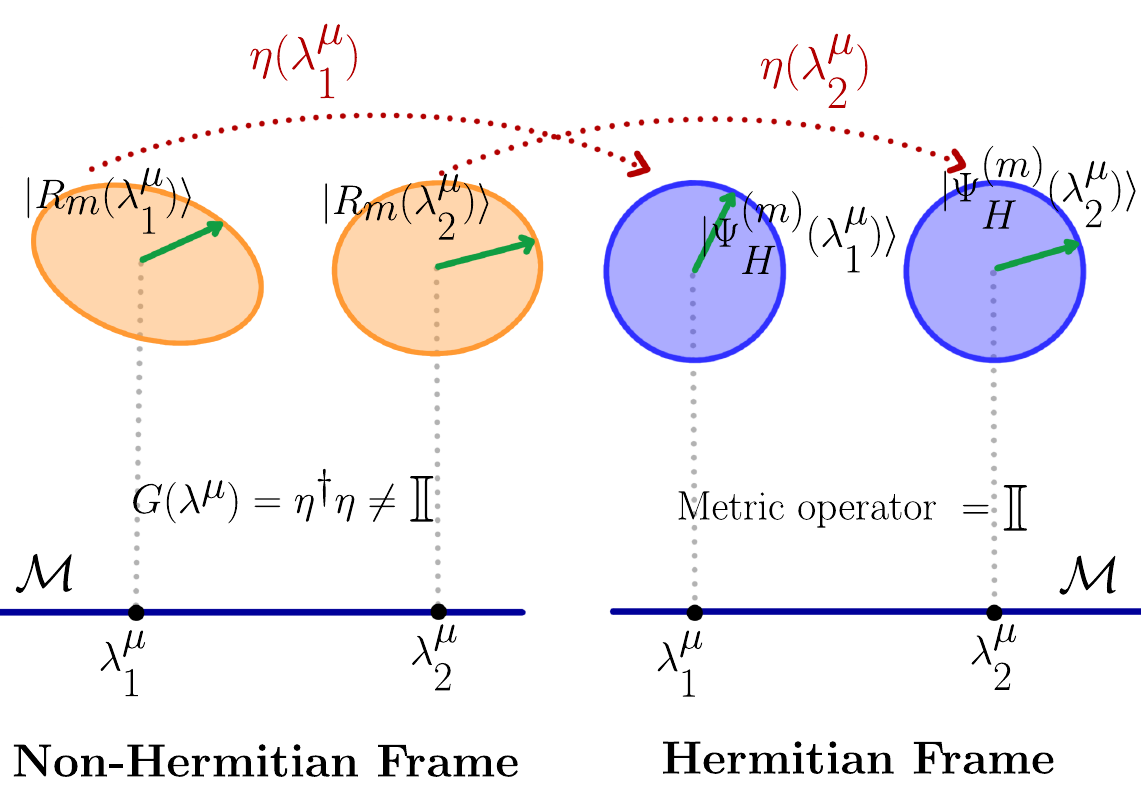}
    \caption{
Schematic illustration of the Dyson map connecting the non-Hermitian (deformed) and Hermitian (undeformed) Hilbert space bundles over parameter space $\mathcal{M}$. At each point $\lambda^\mu$, the right eigenstates $\ket{R_m(\lambda^\mu)}$ span a fiber with a non-trivial metric operator $G(\lambda^\mu)=\eta^\dagger\eta \neq \mathbb{I}$ (left), while the mapped states $\ket{\Psi_H^{(m)}(\lambda^\mu)}=\eta(\lambda^\mu)\ket{R_m(\lambda^\mu)}$ form an orthonormal basis with a trivial metric operator $\mathbb{I}$ (right). The Dyson map $\eta(\lambda^\mu)$ provides a fiber-wise transformation that flattens the Hilbert space geometry while preserving the spectrum.
}
    \label{fig:dyson_map}
\end{figure}

%
Consider an $N$-dimensional non-Hermitian Hamiltonian $\mathcal{H}(\lambda^{\mu})$ depending on a set of real parameters $\lambda^{\mu} \equiv (\lambda^1, \lambda^2, \ldots)$ that take values in a parameter space $\mathcal{M}$. The parameters could represent coupling constants, external fields, or any continuous knobs that deform the system. Throughout this work, we restrict attention to the \emph{quasi-Hermitian regime}: the region of $\mathcal{M}$ in which $\mathcal{H}$ admits a positive-definite metric operator, real spectrums and the eigenstates form a complete biorthogonal basis. This regime is bounded by the EPs, where eigenvalues and eigenstates coalesce.
Since $\mathcal{H} \neq \mathcal{H}^\dagger$, its left and right eigenstates are distinct~\cite{brody2014biorthogonal, moiseyev2011non}. The right eigenstates $\ket{R_m}$ and left eigenstates $\bra{L_m}$ satisfy
\begin{equation}\label{eq:LR_eigen}
    \mathcal{H}\ket{R_m} = E_m\ket{R_m},
    \qquad
    \bra{L_m}\mathcal{H}^\dagger = E_m\bra{L_m},
\end{equation}
and obey the biorthogonality condition $\braket{L_n}{R_m} = \delta_{nm}$ rather than the standard orthonormality. This is the defining feature of the non-Hermitian Hilbert space, the inner product structure is deformed, and the geometry of the state space is no longer the familiar one. 
Over each point $\lambda^{\mu} \in \mathcal{M}$, the states $\ket{R_m(\lambda^\mu)}$ span a fiber of a bundle $\mathcal{E}_{\mathcal{H}} \to \mathcal{M}$. In each such fiber, the inner product is biorthogonal, such that the metric operator $G(\lambda^\mu)$ on the fiber is not the identity but is encoded in the relation between left and right eigenstates, where:
\begin{align}
G &= \sum_{i=1}^{N} |L_i\rangle \langle L_i|, 
\quad 
G^{-1} = \sum_{i=1}^{N} |R_i\rangle \langle R_i|, \\
\mathbb{I} &= \sum_{i=1}^{N} |R_i\rangle \langle L_i| .
\label{eq:G_lambda}
\end{align}
Because this metric operator $G$ varies from fiber to fiber as the parameters change, the bundle $\mathcal{E}_{\mathcal{H}}$ carries a non-trivial metric structure ($G \neq \mathbb{I}$). We refer to the resulting structure as the \emph{deformed Hilbert space}, or equivalently the \emph{non-Hermitian frame}.
The passage from the non-Hermitian to the Hermitian description is achieved by the \emph{Dyson map} $\eta(\lambda^\mu)$: an invertible, parameter-dependent operator that transforms $\mathcal{H}$ into an equivalent Hermitian Hamiltonian $H(\lambda^\mu)$ through a similarity transformation~\cite{dyson1956general, mostafazadeh2002pseudo, mostafazadeh2002pseudo2, mostafazadeh2002pseudo3},
\begin{equation}\label{eq:similarity}
    H = \eta\,\mathcal{H}\,\eta^{-1},
    \qquad H = H^\dagger.
\end{equation}
The Hamiltonian $H$ has the same eigenvalues as $\mathcal{H}$, but it is Hermitian with respect to the standard inner product. The Dyson map acts on the states as $\ket{\Psi_H^{(m)}} = \eta\ket{R_m}$, mapping each right eigenstate of $\mathcal{H}$ to an eigenstate of $H$ with the same eigenvalue, i.e. $H\ket{\Psi_H^{(m)}} = E_m\ket{\Psi_H^{(m)}}$. The mapped states are orthonormal under the standard inner product:
\begin{align}
\braket{\Psi_H^{(n)}}{\Psi_H^{(m)}}
&= \bra{R_n}\eta^\dagger\eta\ket{R_m}\nonumber \\
&= \bra{R_n}\,G\,\ket{R_m}\nonumber \\
&= \braket{L_n}{R_m} = \delta_{nm}.
\label{eq:orthonormality}
\end{align}
where $G = \eta^\dagger\eta$ and $\bra{R_n}\,G =\bra{L_n}$ using Eq.~\eqref{eq:G_lambda}. Equation~\eqref{eq:orthonormality} reveals the central role of the Dyson map, it absorbs the deformed inner product of the non-Hermitian Hilbert space into the map itself, so that the mapped states live in a Hilbert space with the standard, undeformed inner product. The biorthogonal pairing $\braket{L_n}{R_m}$ in the non-Hermitian Hilbert space becomes the standard orthogonality $\braket{\Psi_H^{(n)}}{\Psi_H^{(m)}}$ in the Hermitian Hilbert space.
The states $\ket{\Psi_H^{(m)}(\lambda^\mu)}$ span a fiber over each point $\lambda^\mu \in \mathcal{M}$, forming the corresponding Hermitian bundle $\mathcal{E}_H \to \mathcal{M}$. In each fiber of this bundle, the states are orthonormal under the standard inner product, and the familiar tools of quantum mechanics (completeness relations, spectral decompositions, orthogonal projections) apply without modification. The metric operator on every fiber is the identity, independent of $\lambda^\mu$, so the bundle $\mathcal{E}_H$ carries a trivial metric structure. We refer to this as the \emph{undeformed Hilbert space}, or equivalently the \emph{Hermitian frame}.  The Dyson map is therefore the transformation that flattens the geometry; it maps the geometry of deformed Hilbert space to the undeformed Hilbert space. Figure~(\ref{fig:dyson_map}) provides a geometric picture of this construction.
The two bundles $\mathcal{E}_{\mathcal{H}}$ and $\mathcal{E}_H$ describe the same physics. They have the
same base space $\mathcal{M}$, the same eigenvalues $E_m$, and the same number of states in each fiber. The only difference is the inner product on the fibers: deformed (biorthogonal) in $\mathcal{E}_{\mathcal{H}}$, standard (orthonormal) in $\mathcal{E}_H$. The Dyson map is the fiber-wise map that intertwines the two,
\begin{equation}\label{eq:bundle_map}
    \eta:\quad
    \mathcal{H}
    \longrightarrow
    H,
    \qquad
    \ket{R_m}
    \longmapsto
    \ket{\Psi_H^{(m)}}.
\end{equation}
It is an isomorphism of vector spaces at each $\lambda^\mu$, but it is not unitary; it changes the inner product, which is precisely its purpose. The Dyson map is not unique. If $\eta$ satisfies $H = \eta \mathcal{H} \eta^{-1}$, then
\begin{equation}
\eta \;\to\; \bar{\eta} = U\,\eta, \quad
\text{with}
~~U^\dagger U = \mathbb{I}
\label{eq:dyson-gauge}
\end{equation}
$\bar{\eta} \mathcal{H} \bar{\eta}^{-1} = \bar{H}$, where $\bar{H} =  U H U^\dagger$. The metric operator is invariant under this transformation $\eta \;\to\; \bar{\eta}$ as $\bar{G} = \bar{\eta}^\dagger \bar{\eta} = \eta^\dagger U^\dagger U \eta = G$, so all the physical content is encoded in $G$, while $\eta$ carries a unitary redundancy known as the \emph{Dyson gauge freedom}. A choice of $\eta(\lambda^\mu)$ corresponds to a choice of gauge, selecting a particular Hermitian frame without affecting physical observables, and only those quantities which are invariant under $\eta \;\to\; \bar{\eta}$ are physically meaningful. The complete structure of this unitary freedom, including the classification of gauge-invariant, gauge-covariant, and gauge-dependent quantities, is worked out in detail in Appendix~\ref{app:gauge_structure}, and all gauge-transformation properties used throughout this work are taken directly from there.\\
As $\lambda^\mu$ varies, the Dyson map changes, and with it the relationship between the deformed and undeformed Hilbert spaces changes; the rate of this change is captured by the \emph{Dyson connection},
\begin{equation}\label{eq:dyson_connection}
    \Gamma_\mu \equiv (\partial_\mu \eta)\,\eta^{-1},
\end{equation}
where $\partial_\mu \equiv \frac{\partial}{\partial \lambda^\mu}$. The Dyson connection is an operator-valued one-form on parameter space that plays the role of a gauge connection on the undeformed Hilbert space. Since, $\Gamma_\mu$ is constructed from a single invertible map $\eta$, it is a \emph{pure-gauge} connection and satisfies the zero-curvature condition:
\begin{equation}\label{eq:flatness}
    \mathcal{F}_{\mu\nu}^{(\Gamma)}
    \equiv \partial_\mu\Gamma_\nu
    - \partial_\nu\Gamma_\mu
    - [\Gamma_\mu,\,\Gamma_\nu] = 0.
\end{equation}
We define $\mathcal{F}_{\mu\nu}^{(\Gamma)}$ as the curvature of the Dyson connection or equivalently \emph{$\Gamma$-field strength}. This flatness has a direct geometric meaning: the Dyson map $\eta$ provides a smooth, single-valued transformation from the deformed Hilbert space to the undeformed Hilbert space at every point in parameter space. The vanishing of $\mathcal{F}_{\mu\nu}^{(\Gamma)}$ implies that this flattening is globally consistent, so parallel transport generated by $\Gamma_\mu$ around any closed loop produces no holonomy. As we will see in the next section, non-trivial holonomy (and hence the non-Hermitian Berry phase) arises not from $\Gamma_\mu$ itself, but from its \emph{projected} components when restricted to individual energy eigenstates.\\
In this work, all geometric constructions (connections, curvatures, the Berry phase, the QGT) are carried out in the Hermitian frame, where the inner product is the standard one and the tools of the usual quantum geometry apply directly. The non-Hermitian origin of the system is not lost: it is encoded entirely in the parameter dependence of $\eta$. The decomposition of dyson connection is the subject of the next section. Also, throughout the next few sections, we work with a fixed (target) eigenstate $\ket{\Psi_H^{(m)}}$ and suppress the index $(m)$ where no ambiguity arises.
%
\section{Decomposition of the Dyson Connection}
\label{sec:decomposition}
%
The Dyson connection $\Gamma_\mu$ [Eq.~\eqref{eq:dyson_connection}] encodes how the Dyson map $\eta$ varies across parameter space. Since any operator can be uniquely decomposed into Hermitian and anti-Hermitian parts, applying this to $\Gamma_\mu$ yields
\begin{equation}\label{eq:Gamma_decomp}
    \Gamma_\mu = S_\mu + K_\mu,
\end{equation}
with
\begin{align}
    S_\mu &\equiv \tfrac{1}{2}(\Gamma_\mu + \Gamma_\mu^\dagger),
    \qquad S_\mu^\dagger = S_\mu,
    \label{eq:S_def}\\[4pt]
    K_\mu &\equiv \tfrac{1}{2}(\Gamma_\mu - \Gamma_\mu^\dagger),
    \qquad K_\mu^\dagger = -K_\mu.
    \label{eq:K_def}
\end{align}
To uncover the geometric content of $S_\mu$ and $K_\mu$, we examine their action on the norm of a state under the infinitesimal transformation
$\ket{\Psi_H} \mapsto
(\mathbb{I}+S_\mu\,d\lambda^\mu)\ket{\Psi_H}$:
\begin{equation}\label{eq:S_norm}
    \|(\mathbb{I}+S_\mu\,d\lambda^\mu)\ket{\Psi_H}\|^2
    = 1 + 2\,\bra{\Psi_H}S_\mu\ket{\Psi_H}\,d\lambda^\mu
      + \mathcal{O}(d\lambda^2),
\end{equation}
while for $K_\mu$,
\begin{equation}\label{eq:K_norm}
\begin{split}
    \|(\mathbb{I}+K_\mu\,d\lambda^\mu)\ket{\Psi_H}\|^2
   & = 1 + \bra{\Psi_H}
      (K_\mu + K_\mu^\dagger)
      \ket{\Psi_H}\,d\lambda^\mu\\
     & + \mathcal{O}(d\lambda^2)\\
    &= 1 + \mathcal{O}(d\lambda^2),
    \end{split}
\end{equation}
where $\|\cdot\|$ denotes the standard Hilbert space norm.
Thus $S_\mu$ changes the norm at first order, whereas $K_\mu$ preserves it. Since $S_\mu$ is Hermitian, the coefficient $\bra{\Psi_H}S_\mu\ket{\Psi_H}$ is real; $K_\mu$ is anti-Hermitian, the coefficient $\bra{\Psi_H}K_\mu\ket{\Psi_H}$ is purely imaginary. We collect these into the real projected quantities
\begin{equation}\label{eq:A_S}
    \bra{\Psi_H}S_\mu\ket{\Psi_H} \;\equiv\; \mathcal{A}_\mu^{(S)},
    \qquad
    -i\,\bra{\Psi_H}K_\mu\ket{\Psi_H} \;\equiv\; \mathcal{A}_\mu^{(K)},
\end{equation}
which capture the state-dependent contributions of $S_\mu$ and $K_\mu$. The origin of the norm-changing behaviour of $S_\mu$ is seen by differentiating the metric $G$ with respect to $\lambda^\mu$:
\begin{equation}\label{eq:dG}
    \partial_\mu G
    = \eta^\dagger
      (\Gamma_\mu^\dagger + \Gamma_\mu)\,\eta
    = 2\,\eta^\dagger S_\mu\,\eta,
\end{equation}
where the anti-Hermitian part drops out identically since $K_\mu^\dagger + K_\mu = 0$. Multiplying above relation by $G^{-1} = \eta^{-1}(\eta^\dagger)^{-1}$ from the left yields
\begin{equation}\label{eq:metric_connection}
    \tfrac{1}{2}\,G^{-1}\partial_\mu G
    \;=\;
    \eta^{-1}\,S_\mu\,\eta.
\end{equation}
We define $C_\mu \;\equiv\; \tfrac{1}{2}\,G^{-1}\partial_\mu G$ as the \emph{metric deformation operator}. Equation~\eqref{eq:metric_connection} establishes that $S_\mu$ encodes the information of the metric deformation operator $C_\mu$, mapped from the non-Hermitian to the Hermitian frame by a similarity transformation with the Dyson map. The remaining component $K_\mu = \Gamma_\mu - S_\mu$ therefore encodes the part of the variation of $\eta$ that leaves the metric invariant. This is precisely the infinitesimal form of the unitary freedom, Eq.~\eqref{eq:dyson-gauge}, which preserves $G$ but changes how the non-Hermitian states are embedded into the Hermitian frame. Motivated by these identifications, we refer to $S_\mu$ as the \emph{stretching component} and to $K_\mu$ as the \emph{rotation component} of the Dyson connection, with $\mathcal{A}_\mu^{(S)}$ and $\mathcal{A}_\mu^{(K)}$ denoting the corresponding \emph{stretching} and \emph{rotation projections}, respectively.
%
\subsection{Action of the Stretching and Rotation Components on a Single State}
\begin{figure}[t]
    \centering
    \includegraphics[width=0.9\linewidth]{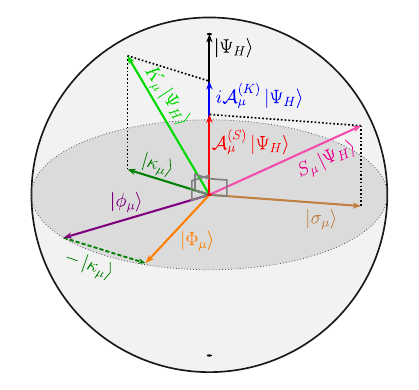}
    \caption{Schematic illustration of the geometric decomposition of the Dyson connection (stretching component ($S_\mu$) and rotation component ($K_\mu$)) acting on a quantum state $\ket{\Psi_H}$ in Hilbert space. $S_\mu$ induces a norm-changing parallel component $\mathcal{A}_\mu^{(S)}\ket{\Psi_H}$ and an orthogonal deformation $\ket{\sigma_\mu}$, while $K_\mu$ generates a phase rotation and an orthogonal mixing $\ket{\kappa_\mu}$. The projected and dressed derivatives, $\ket{\phi_\mu}$ and $\ket{\Phi_\mu}$, are also shown.}
    \label{fig:action}
\end{figure}

%
To extract the physical content of the stretching component $S_\mu$, we examine its action on a fixed eigenstate $\ket{\Psi_H}$ of the Hermitian Hamiltonian $H$.  Using the projectors $\mathcal{P} = \ket{\Psi_H}\bra{\Psi_H}$ and $\mathcal{Q} = \mathbb{I} - \mathcal{P}$, we decompose
\begin{equation}
    S_\mu \ket{\Psi_H}
    = \mathcal{P} S_\mu \ket{\Psi_H}
    + \mathcal{Q} S_\mu \ket{\Psi_H}.
\end{equation}
The parallel component is
\begin{equation}
    \mathcal{P} S_\mu\ket{\Psi_H}
    = \ket{\Psi_H}\bra{\Psi_H}S_\mu\ket{\Psi_H}
    = \mathcal{A}_\mu^{(S)}\,\ket{\Psi_H}.
\end{equation}
The orthogonal component is defined as \emph{stretching fluctuation state}
\begin{equation}\label{eq:sigma}
    \mathcal{Q} S_\mu \ket{\Psi_H} \equiv \ket{\sigma_\mu},
    \quad \text{where} ~
    \braket{\Psi_H}{\sigma_\mu} = 0.
\end{equation}
Thus, the action of $S_\mu$ takes the form
\begin{equation}
    S_\mu \ket{\Psi_H}
    = \mathcal{A}_\mu^{(S)} \ket{\Psi_H} + \ket{\sigma_\mu}.
\end{equation}
The parallel component $\mathcal{A}_\mu^{(S)}\,\ket{\Psi_H}$ acts along the state itself and therefore changes its norm. In contrast, the orthogonal component $\ket{\sigma_\mu}$ tilts the state into the subspace spanned by the remaining eigenstates. The action of $S_\mu$ thus simultaneously rescales the state and changes its direction within Hilbert space. This geometric action is illustrated schematically in Fig.~\ref{fig:action}. To quantify the total deformation, we define the \emph{stretching scalar}
\begin{equation}\label{eq:stretching_scalar}
    \mathcal{I}_\mu 
    \equiv \mel{\Psi_H}{S_\mu^2}{\Psi_H}
    = \bigl(\mathcal{A}_\mu^{(S)}\bigr)^2 + \|\sigma_\mu\|^2,
\end{equation}
where ``scalar'' refers to their being expectation values (scalars in the Hilbert space) labelled by the parameter direction $\mu$. In Eq.~\eqref{eq:stretching_scalar}, both contributions are non-negative: the first measures the stretching along $\ket{\Psi_H}$, while the second measures the deformation into orthogonal directions. The scalar $\mathcal{I}_\mu$ therefore captures the total magnitude of metric deformation along the parameter direction $\mu$. Under a Dyson gauge freedom, $\mathcal{A}_\mu^{(S)}$ and $\mathcal{I}_\mu$ are gauge-invariant, and $\ket{\sigma_\mu}$ transforms covariantly [see appendix~\ref{app:gauge_structure}].\\
%
%
We now consider the rotation component $K_\mu$. Its action on $\ket{\Psi_H}$ can be decomposed in the same way,
\begin{equation}
    K_\mu \ket{\Psi_H}
    = i\,\mathcal{A}_\mu^{(K)} \ket{\Psi_H}
    + \ket{\kappa_\mu},
    \qquad
    \braket{\Psi_H}{\kappa_\mu} = 0.
\end{equation}
Here $\mathcal{A}_\mu^{(K)}$ is rotation projection defined in Eq.~\eqref{eq:A_S}, and $\ket{\kappa_\mu}$ is the \emph{rotation fluctuation state}, defined as
\begin{equation}\label{eq:kappa}
    \ket{\kappa_\mu} \equiv \mathcal{Q} K_\mu \ket{\Psi_H}.
\end{equation}
In contrast to $S_\mu$, the parallel component of $K_\mu$ generates a phase rotation without changing the norm. The orthogonal component $\ket{\kappa_\mu}$ mixes the state with other eigenstates, encoding the unitary rotation of state generated by $K_\mu$, arising from the Dyson gauge freedom. We define the corresponding \emph{rotation scalar}
\begin{equation}\label{eq:j_scalar}
    \mathcal{J}_\mu 
    \equiv -\mel{\Psi_H}{K_\mu^2}{\Psi_H}
    = \bigl(\mathcal{A}_\mu^{(K)}\bigr)^2 + \|\kappa_\mu\|^2.
\end{equation}
This scalar measures the total magnitude of unitary rotation along the parameter direction $\mu$. Unlike $\mathcal{I}_\mu$, the quantity $\mathcal{J}_\mu$ is not gauge-invariant, reflecting the inhomogeneous transformation of $K_\mu$ under Dyson gauge freedom [see appendix~\ref{app:gauge_structure}].
%
\subsection{Dressed Derivative State}
%
We now examine how the eigenstate $\ket{\Psi_H}$ varies with the parameters $\lambda^\mu$. It can be decomposed in parallel and orthogonal components. We define the orthogonal component as
\begin{equation}\label{eq:orthostate}
    \ket{\phi_\mu} \equiv \mathcal{Q}\ket{\partial_\mu \Psi_H},
    \quad \text{with}\quad
    \braket{\Psi_H}{\phi_\mu} = 0,
\end{equation}
which captures the change of $\ket{\Psi_H}$ in directions orthogonal to itself. However, it is not gauge-covariant under a Dyson gauge freedom; it acquires an additional gauge-dependent contribution. A gauge covariant object can be constructed by combining $\ket{\phi_\mu}$ with the rotation fluctuation state $\ket{\kappa_\mu}$. We define the \emph{dressed derivative state} as
\begin{equation}\label{eq:Phi}
    \ket{\Phi_\mu}
    \equiv \ket{\phi_\mu} - \ket{\kappa_\mu}
    = \mathcal{Q}D_\mu^{(K)}\ket{\Psi_H},
\end{equation}
where $D_\mu^{(K)}$ is called \emph{$K$-covariant derivative} and defined as
\begin{equation}\label{eq:K_cov}
    D_\mu^{(K)} \equiv \partial_\mu - K_\mu.
\end{equation}
Both $\ket{\phi_\mu}$ and $\ket{\kappa_\mu}$ transform inhomogeneously under Dyson gauge freedom, acquiring the same gauge-dependent contribution. Their difference therefore eliminates this artifact, and the dressed derivative state transforms covariantly [Eq.~\eqref{eq:app_phi}]. We thus obtain two covariant projected states,
\begin{equation}
    \ket{\Phi_\mu} = \mathcal{Q} D_\mu^{(K)} \ket{\Psi_H},
    \qquad
    \ket{\sigma_\mu} = \mathcal{Q} S_\mu \ket{\Psi_H},
\end{equation}
which transform identically under Dyson gauge freedom. Their inner products are therefore gauge-invariant and form the basic building blocks of the non-Hermitian QGT, which will be discussed in sec.~(\ref{sec:qgt}). In the Hermitian limit, $S_\mu = K_\mu = 0$, and hence $\ket{\Phi_\mu}$ reduces to $\ket{\phi_\mu}$, recovering the usual geometry.
%
\section{The Complex Berry Phase and Berry Curvature}
\label{sec:complex_berry_phase}
%
We now construct the geometric phase associated with adiabatic transport in parameter space, built on the decomposition of the Dyson connection $\Gamma_\mu = S_\mu + K_\mu$ introduced in Sec.~\ref{sec:decomposition}.
\subsection{Non-Hermitian Berry Connection}
When the eigenstate $\ket{\Psi_H(\lambda^\mu)}$ of the Hermitian Hamiltonian $H(\lambda^\mu)$ is transported adiabatically along a path in parameter space, its variation is governed by the \emph{covariant derivative} defined as
\begin{equation}\label{eq:full_cov}
  D_\mu \equiv \partial_\mu - \Gamma_\mu = \partial_\mu - (S_\mu + K_\mu), 
\end{equation}
which incorporates both the intrinsic change of the state and the variation of the Dyson map. Projecting $D_\mu \ket{\Psi_H}$ onto $\ket{\Psi_H}$, and using Eqs.~\eqref{eq:Gamma_decomp} and \eqref{eq:A_S} of Sec.~\ref{sec:decomposition}, we obtain
\begin{equation}\label{eq:D_projection}
\mel{\Psi_H}{D_\mu}{\Psi_H}
= \braket{\Psi_H}{\partial_\mu \Psi_H}
- \mathcal{A}_\mu^{(S)}
- i\,\mathcal{A}_\mu^{(K)}.
\end{equation}
The orthogonal components do not contribute to this projection. The first term on right hand side of Eq.~\eqref{eq:D_projection} defines the \emph{standard Berry connection} in the Hermitian frame,
\begin{equation}\label{eq:berry_connection}
\mathcal{A}_\mu^{(B)} \equiv i\,\braket{\Psi_H}{\partial_\mu \Psi_H},
\end{equation}
which is real as a consequence of normalization of state. Substituting this expression in Eq.~\eqref{eq:D_projection}, the projected covariant derivative becomes
\begin{equation}
\mel{\Psi_H}{D_\mu}{\Psi_H}
= -i\bigl(\mathcal{A}_\mu^{(B)} + \mathcal{A}_\mu^{(K)}\bigr)
- \mathcal{A}_\mu^{(S)}.
\end{equation}
The combination $\mathcal{A}_\mu^{(B)} + \mathcal{A}_\mu^{(K)}$ generates a phase rotation of the state, with $\mathcal{A}_\mu^{(B)}$ arising from the intrinsic parameter dependence of the eigenstate and $\mathcal{A}_\mu^{(K)}$ providing an additional contribution from the Dyson connection. This structure mirrors that of the dressed derivative state $\ket{\Phi_\mu}=\ket{\phi_\mu}-\ket{\kappa_\mu}$: both $\ket{\phi_\mu}$ and $\ket{\kappa_\mu}$ acquire the same gauge artifact, so their difference transforms covariantly [Eq.~\eqref{eq:app_phi}]. An analogous cancellation occurs at the level of the connections, where $\mathcal{A}_\mu^{(B)}$ and $\mathcal{A}_\mu^{(K)}$ are each gauge-dependent but their sum is gauge-invariant and therefore constitutes a physically meaningful quantity. The term $\mathcal{A}_\mu^{(S)}$ is associated with the stretching component $S_\mu$ and reflects the variation of the metric operator, as mentioned in Eq.~\eqref{eq:metric_connection}. It therefore represents a genuine geometric effect, producing a change in the norm of the state. We define the \emph{Non-Hermitian Berry connection} as
\begin{equation}
\mathcal{A}_\mu\equiv i\mel{\Psi_H}{D_\mu}{\Psi_H}
= \bigl(\mathcal{A}_\mu^{(B)} + \mathcal{A}_\mu^{(K)}\bigr)
- i\,\mathcal{A}_\mu^{(S)}.
\label{NHconnection}
\end{equation}
The real part of this connection describes the total phase rotation, while the imaginary part describes the metric deformation. Under a Dyson gauge freedom, the full connection $\mathcal{A}_\mu$ is gauge-invariant as $\mathcal{A_\mu^{(S)}}$ is gauge-invariant [see Eq.~\eqref{eq:app_amu_gauge}].
\subsection{Non-Hermitian Berry Phase}
The \emph{Non-Hermitian Berry phase} accumulated over a closed loop $\mathcal{C}$ in parameter space is given as
\begin{equation}
\gamma = \oint_\mathcal{C} \mathcal{A}_\mu\, d\lambda^\mu.
\end{equation}
Using the definition of the connection~\eqref{NHconnection}, this can be written as
\begin{equation}\label{eq:cbp}
\gamma
= \oint_\mathcal{C}
\bigl(\mathcal{A}_\mu^{(B)} + \mathcal{A}_\mu^{(K)}\bigr)\, d\lambda^\mu
- i \oint_\mathcal{C} \mathcal{A}_\mu^{(S)}\, d\lambda^\mu.
\end{equation}
The real part of $\gamma$ gives the total geometric phase accumulated along the loop, determined by the phase component of the non-Hermitian connection. The imaginary part is determined entirely by $\mathcal{A}_\mu^{(S)}$ and measures the cumulative effect of metric deformation along the loop. It describes how the norm of the state changes after one cycle. Using Eq.~\eqref{eq:cbp}, we get:
\begin{equation}\label{eq:norm_ratio}
\frac{\|\Psi_{\mathrm{final}}\|}{\|\Psi_{\mathrm{initial}}\|}
= e^{\mathrm{Im}\,\gamma} = \exp\!\left(-\oint_\mathcal{C} \mathcal{A}_\mu^{(S)}\, d\lambda^\mu\right).
\end{equation}
Thus, the imaginary part of the complex Berry phase encodes a geometric effect that is absent in Hermitian systems. A nonzero $\mathrm{Im}\,\gamma$ leads to amplification or attenuation of the state upon completing the loop, reflecting the underlying deformation of the metric along the path. Such imaginary contributions to the geometric phase in non-Hermitian systems have been extensively studied and are known to produce observable gain and loss effects~\cite{garrison1988complex, longhi2023complex, longhi2009bloch}. In the Hermitian limit, where $S_\mu = 0$ and $K_\mu = 0$, the Non-Hermitian Berry phase reduces to the standard Berry phase.
\subsection{Non-Hermitian Berry Curvature}
The non-Hermitian Berry phase can be expressed as a surface integral via Stokes' theorem, where the integrand is the corresponding Berry curvature. We first define three curvatures:
\begin{subequations}\label{eq:curvatures}
	\begin{align}
		\Omega_{\mu\nu}^{(B)} &\equiv \partial_\mu\mathcal{A}_\nu^{(B)} 
		- \partial_\nu\mathcal{A}_\mu^{(B)}, \label{eq:omega} \\[4pt]
		\Omega_{\mu\nu}^{(K)} &\equiv \partial_\mu\mathcal{A}_\nu^{(K)} 
		- \partial_\nu\mathcal{A}_\mu^{(K)}, \label{eq:FK} \\[4pt]
		\Omega_{\mu\nu}^{(S)} &\equiv \partial_\mu\mathcal{A}_\nu^{(S)} 
		- \partial_\nu\mathcal{A}_\mu^{(S)}. \label{eq:FS}
	\end{align}
\end{subequations}
Here $\Omega_{\mu\nu}^{(B)}$ is the \emph{standard Berry curvature}, determined by the intrinsic parameter dependence of the eigenstate. The additional curvatures $\Omega_{\mu\nu}^{(K)}$ and $\Omega_{\mu\nu}^{(S)}$ arise from the projected components of the Dyson connection and encode the geometric response associated with the rotation and stretching sectors, respectively. While the connections $\mathcal{A}_\mu^{(B)} + \mathcal{A}_\mu^{(K)}$ and $\mathcal{A}_\mu^{(S)}$ characterize the geometric variation along a path, their curls quantify the corresponding response over a finite area in parameter space. The \emph{non-Hermitian Berry curvature} is written using Eq.~\eqref{NHconnection} as
\begin{align}\label{eq:complex_curvature}
    \Omega_{\mu\nu}
    &\equiv \partial_\mu\mathcal{A}_\nu
          - \partial_\nu\mathcal{A}_\mu \nonumber\\
    &= \bigl(\Omega_{\mu\nu}^{(B)}
       + \Omega_{\mu\nu}^{(K)}\bigr)
       - i\,\Omega_{\mu\nu}^{(S)},
\end{align}
which combines these contributions. The curvatures $\Omega_{\mu\nu}^{(B)}$ and $\Omega_{\mu\nu}^{(K)}$ exhibit the same gauge structure as the connections $\mathcal{A}_\mu^{(B)}$ and $\mathcal{A}_\mu^{(K)}$: neither is individually gauge-invariant, but their sum $\Omega_{\mu\nu}^{(B)}+\Omega_{\mu\nu}^{(K)}$ is. The stretching curvature $\Omega_{\mu\nu}^{(S)}$ is gauge-invariant on its own, mirroring $\mathcal{A}_\mu^{(S)}$. Consequently, the non-Hermitian Berry curvature $\Omega_{\mu\nu}$ is gauge-invariant. By Stokes' theorem, the Non-Hermitian Berry phase over a closed loop $\mathcal{C} = \partial \Sigma$ can be expressed as
\begin{equation}
\gamma
= \int_\Sigma \Omega_{\mu\nu}
\, d\lambda^\mu \wedge d\lambda^\nu,
\end{equation}
where $\Sigma$ is a surface bounded by $\mathcal{C}$ and $\wedge$ denotes the wedge product. The real part of the phase is determined by the flux of $\Omega_{\mu\nu}^{(B)} + \Omega_{\mu\nu}^{(K)}$, while the imaginary part is determined by the flux of $\Omega_{\mu\nu}^{(S)}$.
The curvatures defined above describe the geometric response obtained from projection onto a single eigenstate. However, as shown in Sec.~\ref{sec:decomposition}, the action of $S_\mu$ and $K_\mu$ also generates orthogonal components, $\ket{\sigma_\mu}$ and $\ket{\Phi_\mu}$, which describe how the state deforms and mixes with the rest of the Hilbert space. These contributions are not contained in the non-Hermitian Berry curvature. A complete description of the geometry therefore requires incorporating both the projected response and the orthogonal structure of the state space. This is achieved by the QGT, which unifies the Berry curvature and the QMT within a single object and will be discussed in the subsequent section.
%
\section{The Non-Hermitian QGT}
\label{sec:qgt}
%
The Berry phase and the associated curvature describe the geometric response of a single eigenstate under adiabatic transport in parameter space. However, this description is based on projection onto the state itself and therefore does not capture the contributions from $\ket{\Phi_\mu}$ and $\ket{\sigma_\mu}$ [Sec.~\ref{sec:decomposition}], which encode the intrinsic variation of the eigenstate and the metric deformation respectively. To incorporate these contributions, we introduce the QGT in the biorthogonal (LR) form~\cite{chen2024quantum, hu2025quantum}; the motivation for this choice is discussed in Appendix~\ref{app:RRLL}. The QGT is defined as
\begin{equation}\label{eq:QGT_biorthogonal}
    Q_{\mu\nu}
    = \bra{\partial_\mu L}
      \bigl(\mathbb{I} - \ket{R}\bra{L}\bigr)
      \ket{\partial_\nu R}.
\end{equation}
Using $\ket{R}=\eta^{-1}\ket{\Psi_H}$, $\bra{L}=\bra{\Psi_H}\eta$ and Eq.~\eqref{eq:dyson_connection}, together with $\partial_\mu(\eta\eta^{-1})=0$, this can be written in the Hermitian frame as
\begin{equation}
    Q_{\mu\nu}
    = \bigl(\bra{\partial_\mu\Psi_H}
            + \bra{\Psi_H}\Gamma_\mu\bigr)
      \mathcal{Q}
      \bigl(\ket{\partial_\nu\Psi_H}
            - \Gamma_\nu\ket{\Psi_H}\bigr).
\end{equation}
Using Eqs.~\eqref{eq:Gamma_decomp}-\eqref{eq:K_def},\eqref{eq:sigma},\eqref{eq:kappa}, and \eqref{eq:orthostate} defined in Sec.~\ref{sec:decomposition} and using the property of projection operator $\mathcal{Q}^2=\mathcal{Q}$, the QGT is expanded compactly as
\begin{equation}
    Q_{\mu\nu}
    = \sum_{i,j=1}^{3}
      \alpha^{ij}\,
      \braket{\Psi_\mu^i}{\Psi_\nu^j},
\end{equation}
where $\{\ket{\Psi_\mu^1},\ket{\Psi_\mu^2},\ket{\Psi_\mu^3}\} = \{\ket{\phi_\mu},\ket{\kappa_\mu},\ket{\sigma_\mu}\}$, and the coefficient matrix is
\begin{equation}
    \alpha
    = \begin{pmatrix}
        +1 & -1 & -1 \\
        -1 & +1 & +1 \\
        +1 & -1 & -1
      \end{pmatrix}.
\end{equation}
Substituting $\ket{\phi_\mu}=\ket{\Phi_\mu}+\ket{\kappa_\mu}$ [Eq.~\eqref{eq:Phi}] into the above expression, the $\ket{\kappa_\mu}$ contributions cancel identically and the QGT reduces to the covariant basis $\{\ket{\Phi_\mu},\ket{\sigma_\mu}\}$,
\begin{equation}\label{eq:qgt_eq}
    Q_{\mu\nu}
    = \sum_{a,b=1}^{2}
      \beta^{ab}\,
      \braket{\Psi_\mu^a}{\Psi_\nu^b},
    \quad
    (\ket{\Psi_\mu^1},\ket{\Psi_\mu^2})
    = (\ket{\Phi_\mu},\ket{\sigma_\mu}),
\end{equation}
with
\begin{equation}
    \beta
    = \begin{pmatrix}
        +1 & -1 \\
        +1 & -1
      \end{pmatrix}.
\end{equation}
The structure of the QGT is therefore governed by this $2\times2$ coefficient matrix. Decomposing it into symmetric and antisymmetric parts, one finds
\begin{align}
    \beta_{\mathrm{sym}}  &\equiv \tfrac{1}{2}(\beta+\beta^{T}) = \sigma_z, \\
    \beta_{\mathrm{asym}} &\equiv \tfrac{1}{2}(\beta-\beta^{T}) = -i\sigma_y.
\end{align}
where $\boldsymbol{\sigma} = (\sigma_x,\sigma_y,\sigma_z)$ denotes the Pauli matrices acting  in the reduced two-component space, and superscript $T$ denotes matrix transposition. The non-Hermitian QMT and non-Hermitian Berry curvature are written in terms of QGT as~\cite{provost1980riemannian, zhang2019quantum, chen2024quantum},
\begin{align}
    g_{\mu\nu} &= \tfrac{1}{2}(Q_{\mu\nu} + Q_{\nu\mu})\nonumber\\
    &= \sum_{a,b}\Bigl[
       \sigma_z^{ab}\,
       \mathrm{Re}\,\braket{\Psi_\mu^a}{\Psi_\nu^b}
       + \sigma_y^{ab}\,
       \mathrm{Im}\,\braket{\Psi_\mu^a}{\Psi_\nu^b}
       \Bigr], \label{eq:qgt_g} \\
    \Omega_{\mu\nu}
    &= i(Q_{\mu\nu} - Q_{\nu\mu})\nonumber\\
    &= \sum_{a,b}\Bigl[
       2\sigma_y^{ab}\,
       \mathrm{Re}\,\braket{\Psi_\mu^a}{\Psi_\nu^b}
       - 2\sigma_z^{ab}\,
       \mathrm{Im}\,\braket{\Psi_\mu^a}{\Psi_\nu^b}
       \Bigr],\label{eq:qgt_omega}
\end{align}
where, $ Q_{\mu\nu} = g_{\mu\nu} - \tfrac{i}{2}\,\Omega_{\mu\nu}$. Substituting $\ket{\Psi_\mu^1}=\ket{\Phi_\mu}$ and $\ket{\Psi_\mu^2}=\ket{\sigma_\mu}$ yields the explicit components
\begin{align}
    \mathrm{Re}\,g_{\mu\nu}
    &= \mathrm{Re}\,\braket{\Phi_\mu}{\Phi_\nu}
      - \mathrm{Re}\,\braket{\sigma_\mu}{\sigma_\nu}, \label{eq:re_g} \\
    \mathrm{Im}\,g_{\mu\nu}
    &= -\mathrm{Im}\,\braket{\Phi_\mu}{\sigma_\nu}
       + \mathrm{Im}\,\braket{\sigma_\mu}{\Phi_\nu}, \label{eq:im_g} \\
    \mathrm{Re}\,\Omega_{\mu\nu}
    &= -2\,\mathrm{Im}\,\braket{\Phi_\mu}{\Phi_\nu}
      + 2\,\mathrm{Im}\,\braket{\sigma_\mu}{\sigma_\nu},  \label{eq:re_omega} \\
    \mathrm{Im}\,\Omega_{\mu\nu}
    &= -2\,\mathrm{Re}\,\braket{\Phi_\mu}{\sigma_\nu}
       + 2\,\mathrm{Re}\,\braket{\sigma_\mu}{\Phi_\nu}  \label{eq:im_omega}.
\end{align}
%
\subsection{The Non-Hermitian QMT}
The non-Hermitian QMT $g_{\mu\nu}$ is, in general, complex-valued. Its real and imaginary parts, given in Eqs.~\eqref{eq:re_g} and~\eqref{eq:im_g}, encode distinct geometric aspects of how the eigenstate varies with the parameters. The associated line element,
\begin{equation}\label{eq:line_element}
    ds^2 = g_{\mu\nu}\,d\lambda^\mu\,d\lambda^\nu,
\end{equation}
is therefore complex, with
\begin{equation}\label{eq:ds2_split}
    \mathrm{Re}(ds^2)
    = \mathrm{Re}\,g_{\mu\nu}\,d\lambda^\mu\,d\lambda^\nu,
    \quad
    \mathrm{Im}(ds^2)
    = \mathrm{Im}\,g_{\mu\nu}\,d\lambda^\mu\,d\lambda^\nu.
\end{equation}
These two parts arise from different geometric mechanisms and have distinct physical implications. 
The real part of QMT [Eq.~\eqref{eq:re_g}] contains two competing contributions. The first is a positive semi-definite term associated with the intrinsic parameter dependence of the eigenstate, while the second is a negative semi-definite contribution arising from the stretching fluctuation state. The latter appears whenever the stretching component $S_\mu$ has a nontrivial action in orthogonal subspace. This contribution has no analogue in Hermitian quantum mechanics.
The competition between these two terms leads to a characteristic feature of non-Hermitian geometry. When the intrinsic variation dominates, $\mathrm{Re}(ds^2) > 0$, the neighbouring eigenstates remain distinguishable in the usual sense. When the metric deformation dominates, $\mathrm{Re}(ds^2) < 0$, which indicates that the deformation of the inner product structure exceeds the separation of the states.
The transition between these regimes occurs at a surface in parameter space, defined as a \emph{critical (null) surface},
\begin{equation}\label{eq:metric_transition}
    \mathrm{Re}\,\braket{\Phi_\mu}{\Phi_\nu}\,
    d\lambda^\mu d\lambda^\nu
    =
    \mathrm{Re}\,\braket{\sigma_\mu}{\sigma_\nu}\,
    d\lambda^\mu d\lambda^\nu,
\end{equation}
at which $\mathrm{Re}(ds^2)=0$. This surface marks the onset of indefinite geometry and provides a gauge-invariant boundary between a Hermitian-like region ($\mathrm{Re}(ds^2) > 0$) and a regime where non-Hermitian effects dominate ($\mathrm{Re}(ds^2) < 0$). When present, null surface lies within the quasi-Hermitian region and serves as a precursor to the approach of an EP.\\
This transition may be absent in special cases. If the stretching component acts purely along the eigenstate, $S_\mu\ket{\Psi_H} = \mathcal{A}_\mu^{(S)}\ket{\Psi_H}$, then the orthogonal component vanishes, $\ket{\sigma_\mu}=0$, and QMT reduces to the positive semi-definite form governed entirely by the intrinsic variation. Conversely, if the dressed derivative state vanishes, $\ket{\Phi_\mu}=0$, QMT becomes purely negative semi-definite and the parameter dependence is governed entirely by metric deformation.\\
The imaginary part of QMT, Eq.~\eqref{eq:im_g}, is determined entirely by the coupling between the intrinsic variation and the stretching fluctuation state. It reflects a relative phase between these two contributions in the orthogonal subspace. Unlike the real part, it does not contribute to the distinguishability between neighbouring states, but instead serves as a diagnostic of the internal non-Hermitian geometric structure. It vanishes whenever $\ket{\sigma_\mu}=0$.\\
In the Hermitian limit ($S_\mu=0$, $K_\mu=0$), the stretching fluctuation state vanishes, $\ket{\sigma_\mu}=0$, and the dressed derivative reduces to $\ket{\Phi_\mu}=\ket{\phi_\mu}$. The non-Hermitian QMT then becomes purely real and positive semi-definite,
\begin{equation}\label{eq:limit}
    g_{\mu\nu}
    \;\longrightarrow\;
    g_{\mu\nu}^{(B)}
    = \mathrm{Re}\,\braket{\phi_\mu}{\phi_\nu},
\end{equation}
recovering the standard Fubini--Study metric $g_{\mu\nu}^{(B)}$~\cite{cheng2010quantum}.
\subsection{The Non-Hermitian Berry Curvature from QGT}
The non-Hermitian Berry curvature $\Omega_{\mu\nu}$ is obtained from the antisymmetric part of the QGT  and is, in general, complex-valued. Its real and imaginary parts, given in Eqs.~\eqref{eq:re_omega} and~\eqref{eq:im_omega}, encode geometrically distinct aspects of the adiabatic response.\\
The real part, Eq.~\eqref{eq:re_omega}, governs the geometric phase accumulated under adiabatic transport. Like $g_{\mu\nu}$, it receives contributions from both $\ket{\Phi_\mu}$ and $\ket{\sigma_\mu}$, but through their imaginary part of inner products rather than real, and with opposite relative sign. In the Hermitian limit, where $\ket{\sigma_\mu}=0$, it reduces to the standard Berry curvature $\Omega_{\mu\nu}^{(B)}$,
\begin{equation}
 \Omega_{\mu\nu}
    \;\longrightarrow\;
    \Omega_{\mu\nu}^{(B)} = -2\,\mathrm{Im}\,\braket{\phi_\mu}{\phi_\nu},
\end{equation}
recovering the familiar result. The additional contribution present in the non-Hermitian case reflects the influence of metric deformation on the phase structure, analogous to how $\mathrm{Re}\,g_{\mu\nu}$ acquires a competing negative semi-definite term.\\
The imaginary part in Eq.~\eqref{eq:im_omega} has no Hermitian counterpart. It is determined entirely by the cross-coupling between $\ket{\Phi_\mu}$ and $\ket{\sigma_\mu}$, and vanishes whenever $\ket{\sigma_\mu}=0$. It governs the change in the norm of the eigenstate under adiabatic transport around a closed loop. A nonzero value signals that the state is amplified or attenuated around a closed loop, as quantified by Eq.~\eqref{eq:norm_ratio}, reflecting a path-dependent effect intrinsic to non-Hermitian geometry.
\subsection{Gauge-invariant identities of the Berry curvature}
The curvature $\Omega_{\mu\nu}$ obtained from the QGT in
Eqs.~\eqref{eq:re_omega}--\eqref{eq:im_omega} coincides with the parametric curl of the non-Hermitian connection $\mathcal{A}_\mu$ [Eq.~\eqref{eq:complex_curvature}]. Since $\mathcal{A}_\mu$ is a complex scalar one-form, the corresponding curvature is Abelian. The Abelian nature of the curvature arises from the projection onto a single eigenstate. Separating real and imaginary parts of the curvature in Eq.~\eqref{eq:complex_curvature} , one obtains
\begin{equation}\label{eq:Omega_ReIm_Abelian}
	\mathrm{Re}\,\Omega_{\mu\nu}
	= \Omega_{\mu\nu}^{(B)} + \Omega_{\mu\nu}^{(K)},
	\qquad
	\mathrm{Im}\,\Omega_{\mu\nu}
	= -\Omega_{\mu\nu}^{(S)},
\end{equation}
where $\Omega_{\mu\nu}^{(B)}$, $\Omega_{\mu\nu}^{(K)}$, and $\Omega_{\mu\nu}^{(S)}$ are defined in Eqs.~\eqref{eq:omega}--\eqref{eq:FS}. An equivalent expression for the same curvature is provided by the QGT through covariant inner products in Eqs.~\eqref{eq:re_omega} and \eqref{eq:im_omega}. Equating the two representations yields the identities [see Appendix~\ref{app:identities}]
\begin{align}
    \Omega_{\mu\nu}^{(B)} + \Omega_{\mu\nu}^{(K)}
    &= -2\,\mathrm{Im}\,\braket{\Phi_\mu}{\Phi_\nu}
      + 2\,\mathrm{Im}\,\braket{\sigma_\mu}{\sigma_\nu},
    \label{eq:identity_Re} \\[6pt]
    \Omega_{\mu\nu}^{(S)}
    &= 2\,\mathrm{Re}\,\braket{\Phi_\mu}{\sigma_\nu}
       - 2\,\mathrm{Re}\,\braket{\sigma_\mu}{\Phi_\nu}.
    \label{eq:identity_Im}
\end{align}
These identities express the gauge-invariant combinations of the Berry curvature, $\Omega_{\mu\nu}^{(B)}+\Omega_{\mu\nu}^{(K)}$ and $\Omega_{\mu\nu}^{(S)}$, directly in terms of inner products of the covariant states $\ket{\Phi_\mu}$ and $\ket{\sigma_\mu}$. The left-hand side involves derivatives of the connection components, while the right-hand side depends only on the projected structure of the state in the orthogonal subspace.\\
This relation also has a direct physical implication. Since $\mathrm{Im}\,\Omega_{\mu\nu} = -\Omega_{\mu\nu}^{(S)}$ governs the change in the norm of the eigenstate under adiabatic transport, it follows that this effect is entirely determined by the antisymmetric real overlap between $\ket{\Phi_\mu}$ and $\ket{\sigma_\nu}$. In particular, $\Omega_{\mu\nu}^{(S)}$ vanishes when this coupling is symmetric under $\mu \leftrightarrow \nu$.
\subsection{Field Strengths and the Flatness Constraint}
\label{sec:field_str}
The curvatures $\Omega_{\mu\nu}^{(K)}$ and $\Omega_{\mu\nu}^{(S)}$ govern the complex Berry phase and the non-Hermitian QGT. They are the projections of underlying field strengths $\mathcal{F}_{\mu\nu}^{(K)}$ and $\mathcal{F}_{\mu\nu}^{(S)}$ onto a single state.  Examining these operators reveals the fundamental mechanism by which non-Hermiticity generates geometric structure.\\
The rotation component $K_\mu$ defines the covariant derivative $D_\mu^{(K)} = \partial_\mu - K_\mu$ for states, and the adjoint derivative $\mathcal{D}_\mu^{(K)}(\cdot) = \partial_\mu(\cdot) - [K_\mu, \cdot\,]$ for operators. The field strength for $K_\mu$ is the standard non-Abelian curvature,
\begin{equation}\label{eq:K_field_strength}
    \mathcal{F}_{\mu\nu}^{(K)}
    = \partial_\mu K_\nu - \partial_\nu K_\mu - [K_\mu, K_\nu].
\end{equation}
Unlike the curvatures $\Omega_{\mu\nu}^{(K)}$ and $\Omega_{\mu\nu}^{(S)}$, this field strength transforms covariantly under the Dyson gauge freedom. It measures the non-commutativity of the covariant derivatives, expressed as $[D_\mu^{(K)},\,D_\nu^{(K)}] = -\mathcal{F}_{\mu\nu}^{(K)}$. Physically, $\mathcal{F}_{\mu\nu}^{(K)}$ determines whether the unitary rotation generated by $K_\mu$ acquires an intrinsic twist around a closed loop in parameter space. When $\mathcal{F}_{\mu\nu}^{(K)} = 0$, one can choose a gauge where $K_\mu = 0$ locally, making the unitary description globally consistent. When $\mathcal{F}_{\mu\nu}^{(K)} \neq 0$, the rotation is intrinsically twisted and no such gauge exists.
In contrast, the stretching component $S_\mu$ transforms covariantly as an adjoint field. Its corresponding field strength is the antisymmetrized covariant derivative:
\begin{equation}\label{eq:S_field_strength}
    \mathcal{F}_{\mu\nu}^{(S)}
    = \mathcal{D}_\mu^{(K)} S_\nu - \mathcal{D}_\nu^{(K)} S_\mu.
\end{equation}
Notice that there is no $[S_\mu, S_\nu]$ commutator. In standard gauge theory, this commutator is required because the connection transforms inhomogeneously. Since $S_\mu$ already transforms covariantly, the covariant derivative alone is sufficient. Physically, $\mathcal{F}_{\mu\nu}^{(S)}$ measures whether the metric deformation is path-independent. It tests if stretching along $\lambda^\mu$ then $\lambda^\nu$ gives the same result as the reverse order, after accounting for the unitary rotation.
Since the full Dyson connection $\Gamma_\mu$ is pure gauge, it satisfies the zero-curvature condition $\mathcal{F}_{\mu\nu}^{(\Gamma)} = 0$ [Eq.~\eqref{eq:flatness}]. Expanding this condition gives a flatness constraint that links the rotation and stretching sectors:
\begin{equation}\label{eq:flatness_constraint}
    \mathcal{F}_{\mu\nu}^{(K)} + \mathcal{F}_{\mu\nu}^{(S)} = [S_\mu, S_\nu].
\end{equation}
Using the operator identities derived from this flatness condition [Eqs.~\eqref{eq:app_K_curl} and~\eqref{eq:app_S_curl}], the Eq.~\eqref{eq:flatness_constraint} is expressed as:
\begin{align}
    \mathcal{F}_{\mu\nu}^{(K)} &= [S_\mu, S_\nu], \label{eq:FK_result} \\[4pt]
    \mathcal{F}_{\mu\nu}^{(S)} &= 0. \label{eq:FS_result}
\end{align}
These relations encode the full geometric impact of the Dyson map. Equation~\eqref{eq:FS_result} shows that the stretching component is covariantly symmetric ($\mathcal{D}_\mu^{(K)} S_\nu = \mathcal{D}_\nu^{(K)} S_\mu$). This means the metric deformation does not depend on the order of parameters, relative to the unitary rotation generated by $K_\mu$. Although $\mathcal{F}_{\mu\nu}^{(S)}=0$, its projection $\Omega_{\mu\nu}^{(S)}$ onto a single eigenstate can remain nonzero, leaving the Berry phase complex [Eq.~\eqref{eq:cbp}].
%
From Eq.~\eqref{eq:FK_result}, the rotation field strength is generated entirely by the non-commutativity of the stretching components. If the stretching components commute ($[S_\mu, S_\nu] = 0$), the rotation field strength vanishes and the unitary rotation generated by $K_\mu$ can be gauged away; otherwise it acquires an intrinsic twist.\\
This operator-level structure manifests directly in the QGT. If we project the rotation field strength $\mathcal{F}_{\mu\nu}^{(K)} = [S_\mu, S_\nu]$ onto the eigenstate $\ket{\Psi_H}$ yields:
\begin{equation}\label{eq:FK_onshell_projection}
    -i\mel{\Psi_H}{\mathcal{F}_{\mu\nu}^{(K)}}{\Psi_H} = 2\,\mathrm{Im}\braket{\sigma_\mu}{\sigma_\nu}.
\end{equation}
which follows from the identity $\mel{\Psi_H}{[S_\mu, S_\nu]}{\Psi_H} = 2i\,\mathrm{Im}\mel{\Psi_H}{S_\mu S_\nu}{\Psi_H}$ and Eq.~\eqref{eq:app_nine_inner}. This real quantity depends entirely on the stretching fluctuation state $\ket{\sigma_\mu}$. As shown in Eq.~\eqref{eq:re_omega}, this exact term in the right-hand side provides the non-Hermitian contribution to the real Berry curvature. Therefore, the incompatibility of metric deformations along different parameter directions acts as the fundamental source of non-Hermitian geometric structure.
%
\section{Geometry near Exceptional Point}
\label{sec:ep_singularity}
%
EPs define the boundary of the quasi-Hermitian regime. At an EP, the eigenvalues and eigenstates of $\mathcal{H}$ coalesce, leading to a simultaneous closing of the energy gap ($\Delta E \to 0$) and the loss of positive-definiteness of the metric operator ($g_c \to 0$, where $g_c$ denotes the vanishing eigenvalue of $G$). These are spectral and geometric manifestations of the same underlying eigenvector coalescence. Consequently, the fiber-bundle structure over the parameter space $\mathcal{M}$ becomes ill-defined, and the associated geometric quantities diverge.\\
Let $\lambda_*^{\mu} = (\lambda_*^1, \lambda_*^2, \ldots)$ denote the location of the EP in parameter space. To analyze the singular behavior, we define the parameter displacement
\begin{equation}\label{eq:epsilon_def}
    \epsilon_\mu \equiv \lambda^\mu - \lambda_*^\mu, \qquad \mu = 1, 2, \ldots
\end{equation}
and approach the EP along individual parameter axes, taking $\epsilon_\mu \to 0$ at fixed $\epsilon_\nu = 0$ for all $\nu \neq \mu$. The singular behavior of the geometric structure can be traced to the divergence of the covariant states $\ket{\Phi_\mu}$ and $\ket{\sigma_\mu}$, and consequently to the components of the non-Hermitian QGT.

\subsection{Divergence of the Stretching Component}
\label{subsec:Smu_divergence}
The metric operator $G$ is positive definite away from the EP, it admits the spectral decomposition
\begin{equation}
    G = \sum_{n=1}^{N} g_n \ket{e_n}\bra{e_n}, \qquad g_n > 0,
\end{equation}
where $g_n$ and $\ket{e_n}$ denote the eigenvalues and orthonormal eigenvectors of $G$, respectively. At the EP, at least one eigenvalue vanishes. Let $g_c$ denote an eigenvalue such that $g_c \to 0$ as $\epsilon_\mu \to 0$. The matrix elements of the metric deformation operator $C_\mu$ [Eq.~\eqref{eq:metric_connection}] in the eigenbasis of $G$ are
\begin{equation}\label{eq:Cmu_ep}
    \mel{e_p}{C_\mu}{e_l} = \frac{\mel{e_p}{\partial_\mu G}{e_l}}{2 g_p}.
\end{equation}
To analyze the behavior of the numerator near the EP, we differentiate the eigenvalue equation
$G\ket{e_l} = g_l\ket{e_l}$ and project onto
$\bra{e_p}$ with $p \neq l$:
\begin{equation}\label{eq:cl}
    \mel{e_p}{\partial_\mu G}{e_l}
    = (g_l - g_p)\braket{e_p}{\partial_\mu e_l}.
\end{equation}
For the collapsing direction, $p = c$, this yields
\[
\mel{e_c}{\partial_\mu G}{e_l}
= (g_l - g_c)\braket{e_c}{\partial_\mu e_l}
\;\to\;
g_l \braket{e_c}{\partial_\mu e_l}
\quad \text{as } g_c \to 0.
\]
Thus, provided $\braket{e_c}{\partial_\mu e_l}$ remains finite, the numerator remains finite near the EP. The divergent behaviour of $C_\mu$ is therefore controlled entirely by the prefactor $1/g_p$. For the collapsing direction $p = c$, the matrix element diverges:
\begin{equation}\label{eq:Cmu_diverge}
    \mel{e_c}{C_\mu}{e_l} \sim \frac{1}{g_c} \to \infty \quad \text{for any } l \text{ with }
    \mel{e_c}{\partial_\mu G}{e_l} \neq 0.
\end{equation}
As the stretching component $S_\mu$ is related to $C_\mu$ by a similarity transformation [Eq.~\eqref{eq:metric_connection}], it preserves the spectrum and inherits this divergence: $S_\mu \to \infty$ as $\epsilon_\mu \to 0$. The collapsing direction $\ket{e_c}$ of $G$ in the non-Hermitian frame, along which $C_\mu$ diverges, is mapped to $\eta\ket{e_c}$ in the Hermitian frame. This shows that the singularity is faithfully transported by the Dyson map.\\
The stretching projection $\mathcal{A}_\mu^{(S)}$ can be expressed through the spectral representation of $C_\mu$
\begin{align}\label{eq:As_spectral}
    \mathcal{A}_\mu^{(S)}
    &= \mel{\Psi_H^{(m)}}{\eta\,C_\mu\,\eta^{-1}}{\Psi_H^{(m)}} \notag \\
    &= \sum_{p=1}^{N}\sum_{l=1}^{N}
      \frac{\mel{\Psi_H^{(m)}}{\eta}{e_p}
            \mel{e_p}{\partial_\mu G}{e_l}
            \mel{e_l}{\eta^{-1}}{\Psi_H^{(m)}}}{2\,g_p}.
\end{align}
The dominant contribution in the limit $g_c \to 0$ becomes
\begin{equation}\label{eq:As_diverge}
    \mathcal{A}_\mu^{(S)} \approx \frac{1}{2 g_c} \sum_{l=1}^{N} \mel{\Psi_H^{(m)}}{\eta}{e_c} \mel{e_c}{\partial_\mu G}{e_l} \mel{e_l}{\eta^{-1}}{\Psi_H^{(m)}} \to \infty,
\end{equation}
provided $\mel{\Psi_H^{(m)}}{\eta}{e_c} \neq 0$ i.e. divergence occurs generically whenever the target state $\ket{\Psi_H^{(m)}}$ has a non-zero overlap with the collapsing direction $\eta\ket{e_c}$.  The divergence of $\mathcal{A}_\mu^{(S)}$ implies that the norm of the eigenstate is rescaled at an unbounded rate as the EP is approached, leading to a divergence of the imaginary part of the non-Hermitian Berry connection. Consequently, the stretching curvature $\Omega_{\mu\nu}^{(S)}$ also diverges. Since all non-Hermitian geometric structures in this framework are constructed from $S_\mu$ and its projections, its divergence constitutes the origin of the geometric singularity at the EP.
\subsection{Divergence of the covariant states}
\label{subsec:covariant_divergence}
The squared norm of the stretching fluctuation state $\ket{\sigma_\mu}$ admits a spectral decomposition in the orthonormal eigenbasis $\{\ket{\Psi_H^{(n)}}\}$ of $H$:
\begin{equation}\label{eq:sigma_spectral}
    \|\sigma_\mu\|^2
    = \sum_{n \neq m}
      \bigl|\mel{\Psi_H^{(n)}}{S_\mu}{\Psi_H^{(m)}}\bigr|^2.
\end{equation}
The off-diagonal matrix elements of $S_\mu$ are obtained from the parametric evolution of $H$ together with Eq.~\eqref{eq:S_def}, yielding (see Appendix~\ref{app:Smu_offdiag})
\begin{equation}\label{eq:Smu_offdiag}
    \mel{\Psi_H^{(n)}}{S_\mu}{\Psi_H^{(m)}}
    = \frac{\mathcal{M}_{\mu}^{(nm)}}{E_m - E_n},
\end{equation}
with
\begin{equation}\label{eq:transition_amplitude}
    \mathcal{M}_{\mu}^{(nm)}
    \equiv \frac{1}{2}\mel{\Psi_H^{(n)}}{\bigl(
    {\eta^{\dagger}}^{-1}\partial_\mu \mathcal{H}^\dagger\,\eta^\dagger
    - \eta\,\partial_\mu \mathcal{H}\,\eta^{-1}\bigr)}{\Psi_H^{(m)}}.
\end{equation}
These off-diagonal elements are determined by $\eta(\partial_\mu H)\eta^{-1}$. As $g_c \to 0$, the $\eta^{-1}$ matrix becomes unbounded, and these matrix elements $\mathcal{M}_{\mu}^{(nm)}$ generically inherit this singular behavior at the EP. As the target state $\ket{\Psi_H^{(m)}}$ coalesces with a partner state $\ket{\Psi_H^{(c)}}$, the gap $\Delta E = E_m - E_c$ vanishes, and the sum~\eqref{eq:sigma_spectral} is dominated by the term $n = c$:
\begin{equation}\label{eq:sigma_diverge}
    \|\sigma_\mu\|^2
    = \frac{|\mathcal{M}_\mu^{(cm)}|^2}{|\Delta E|^2}
    + \sum_{\substack{n \neq m \\ n \neq c}}
      \frac{|\mathcal{M}_\mu^{(nm)}|^2}{|E_m - E_n|^2},
\end{equation}
where the first term diverges while the remaining sum stays finite. Near the EP, we suppose $\Delta E \sim |\epsilon_\mu|^{\alpha_\mu}$ and $|\mathcal{M}_\mu^{(cm)}| \sim |\epsilon_\mu|^{-\beta_\mu}$ with $\alpha_\mu, \beta_\mu > 0$, where the exponents characterize the scaling along the $\mu$-th parameter direction. Consequently,
\begin{equation}\label{eq:sigma_scaling}
	\|\sigma_\mu\|^2 \sim |\epsilon_\mu|^{-2(\alpha_\mu+\beta_\mu)}.
\end{equation}
For a second-order EP, a single partner state $\ket{\Psi_H^{(c)}}$ coalesces with the target and dominates the sum. For higher-order EPs, multiple partners coalesce and contribute additional divergent terms, but the leading divergence is governed by the dominant pair; the scaling form~\eqref{eq:sigma_scaling} remains valid, with the exponents determined by the EP order.\\
The stretching scalar $\mathcal{I}_\mu$ (using Eqs.~\eqref{eq:stretching_scalar} and~\eqref{eq:sigma_diverge}) takes the form
\begin{equation}\label{eq:I_diverge}
    \mathcal{I}_\mu
    = \bigl|\mathcal{A}_\mu^{(S)}\bigr|^2
    + \frac{|\mathcal{M}_\mu^{(cm)}|^2}{|\Delta E|^2}
    + \sum_{\substack{n \neq m \\ n \neq c}}
      \frac{|\mathcal{M}_\mu^{(nm)}|^2}{|E_m - E_n|^2},
\end{equation}
where the first two terms diverge at the EP. This stretching scalar senses the EP whenever $\ket{\Psi_H^{(m)}}$ has overlap with the collapsing direction. In the nongeneric case where $S_\mu\ket{\Psi_H^{(m)}} = 0$, both $\mathcal{A}_\mu^{(S)}$ and $\ket{\sigma_\mu}$ vanish, yielding $\mathcal{I}_\mu = 0$ despite the divergence of $S_\mu$. The singularity is therefore present in the geometry but invisible to this particular state. A state-independent diagnostic that probes the stretching across the entire Hilbert space is provided by the \emph{system-level scalar} $\operatorname{tr}(S_\mu S_\nu)$. Using the invariance of the trace under similarity transformations, $\operatorname{tr}(S_\mu S_\nu) = \operatorname{tr}(C_\mu C_\nu)$, and expanding it in the eigenbasis of $G$, we obtain
\begin{equation}\label{eq:tr_expansion}
    \operatorname{tr}(C_\mu C_\nu)
    = \sum_{n=1}^{N}\sum_{p=1}^{N}
      \frac{\mel{e_n}{\partial_\mu G}{e_p}\,
            \mel{e_p}{\partial_\nu G}{e_n}}
           {4\,g_n\,g_p}.
\end{equation}
Using Eq.~\eqref{eq:cl}, as $g_c \to 0$, the dominant contribution arises from the collapsing direction $n = p = c$, yielding
\begin{equation}\label{eq:trace_leading}
    \operatorname{tr}(C_\mu C_\nu)
    = \frac{(\partial_\mu g_c)(\partial_\nu g_c)}{4\,g_c^2}
    + \mathcal{O}\left(\frac{1}{g_c}\right),
\end{equation}
which diverges as $\frac{1}{g_c^{2}}$, making it an unconditional diagnostic of the EP.\\
In a manner analogous to the stretching fluctuation state, the matrix elements of the dressed derivative state $\ket{\Phi_\mu}$ take the form
\begin{equation}\label{eq:Phi_matrix_element}
    \mel{\Psi_H^{(n)}}{D_\mu^{(K)}}{\Psi_H^{(m)}}
    = \frac{\mathcal{N}_\mu^{(nm)}}{E_m - E_n},
\end{equation}
where
\begin{equation}\label{eq:N_amplitude}
    \mathcal{N}_{\mu}^{(nm)}
    \equiv \frac{1}{2}\mel{\Psi_H^{(n)}}{\bigl(
    {\eta^{\dagger}}^{-1}\partial_\mu \mathcal{H}^\dagger\,\eta^\dagger
    + \eta\,\partial_\mu \mathcal{H}\,\eta^{-1}\bigr)}{\Psi_H^{(m)}}.
\end{equation}
The behavior of $\mathcal{N}_\mu^{(nm)}$ can differ from that of $\mathcal{M}_\mu^{(nm)}$ and exhibit independent scaling. This difference originates from their distinct algebraic structures. The quantity $\mathcal{M}_\mu^{(nm)}$ is defined through an anti-Hermitian combination, while $\mathcal{N}_\mu^{(nm)}$ arises from a Hermitian combination of the same operators as seen from Eqs.~\eqref{eq:transition_amplitude} and~\eqref{eq:N_amplitude}. We therefore introduce an independent exponent $\delta_\mu$ such that $|\mathcal{N}_\mu^{(cm)}| \sim |\epsilon_\mu|^{-\delta_\mu}$. The squared norm
\begin{equation}\label{eq:Phi_diverge}
	\|\Phi_\mu\|^2
	= \frac{|\mathcal{N}_\mu^{(cm)}|^2}{|\Delta E|^2}
	+ \sum_{\substack{n \neq m \\ n \neq c}}
	\frac{|\mathcal{N}_\mu^{(nm)}|^2}{|E_m - E_n|^2},
\end{equation}
is dominated by the term $n = c$ and scales as
\begin{equation}
    \|\Phi_\mu\|^2 \sim |\epsilon_\mu|^{-2(\alpha_\mu+\delta_\mu)}.
\end{equation}
Thus, both covariant states exhibit distinct scaling, governed by the coalescing pair, while all other terms in the sum with $n \neq c$ remain finite.
\subsection{Divergence of the QGT Components}
\label{subsec:qgt_divergence}
The components of the non-Hermitian QGT are constructed from inner products of the covariant states $\ket{\Phi_\mu}$ and $\ket{\sigma_\mu}$ [Eqs.~\eqref{eq:qgt_g},~\eqref{eq:qgt_omega}]. As the norms of the covariant states $\ket{\sigma_\mu}$ and $\ket{\Phi_\mu}$ generally scale differently near the EP, their inner products are dominated by the contribution from the coalescing pair $n=c$. Isolating this leading term yields:
\begin{align}
    \braket{\Phi_\mu}{\Phi_\nu} &= \frac{\mathcal{N}_\mu^{(cm)*} \mathcal{N}_\nu^{(cm)}}{|\Delta E|^2} + \mathcal{O}(\epsilon^0), \label{eq:PhiPhi_leading} \\
    \braket{\sigma_\mu}{\sigma_\nu} &= \frac{\mathcal{M}_\mu^{(cm)*} \mathcal{M}_\nu^{(cm)}}{|\Delta E|^2} + \mathcal{O}(\epsilon^0), \label{eq:sigsig_leading} \\
    \braket{\Phi_\mu}{\sigma_\nu} &= \frac{\mathcal{N}_\mu^{(cm)*} \mathcal{M}_\nu^{(cm)}}{|\Delta E|^2} + \mathcal{O}(\epsilon^0). \label{eq:Phisig_leading}
\end{align}
Substituting these into the non-Hermitian QGT decomposition [Eqs.~\eqref{eq:re_g}--\eqref{eq:im_omega}] explicitly reveals the leading singular behavior of each geometric tensor component:
\begin{widetext}
	\begin{subequations}\label{eq:qgt_scaling}
		\begin{align}
			\mathrm{Re}\,g_{\mu\nu} &= \frac{1}{|\Delta E|^2} \left[ \mathrm{Re} \left( \mathcal{N}_\mu^{(cm)*} \mathcal{N}_\nu^{(cm)} \right) - \mathrm{Re} \left( \mathcal{M}_\mu^{(cm)*} \mathcal{M}_\nu^{(cm)} \right) \right] + \mathcal{O}(\epsilon^0), \label{eq:Reg_scaling} \\
			\mathrm{Im}\,g_{\mu\nu} &= -\frac{1}{|\Delta E|^2} \left[ \mathrm{Im} \left( \mathcal{N}_\mu^{(cm)*} \mathcal{M}_\nu^{(cm)} \right) - \mathrm{Im} \left( \mathcal{M}_\mu^{(cm)*} \mathcal{N}_\nu^{(cm)} \right) \right] + \mathcal{O}(\epsilon^0), \label{eq:Img_scaling} \\
			\mathrm{Re}\,\Omega_{\mu\nu} &= \frac{2}{|\Delta E|^2} \left[ \mathrm{Im} \left( \mathcal{M}_\mu^{(cm)*} \mathcal{M}_\nu^{(cm)} \right) - \mathrm{Im} \left( \mathcal{N}_\mu^{(cm)*} \mathcal{N}_\nu^{(cm)} \right) \right] + \mathcal{O}(\epsilon^0), \label{eq:ReOmega_scaling} \\
			\mathrm{Im}\,\Omega_{\mu\nu} &= -\frac{2}{|\Delta E|^2} \left[ \mathrm{Re} \left( \mathcal{N}_\mu^{(cm)*} \mathcal{M}_\nu^{(cm)} \right) - \mathrm{Re} \left( \mathcal{M}_\mu^{(cm)*} \mathcal{N}_\nu^{(cm)} \right) \right] + \mathcal{O}(\epsilon^0). \label{eq:ImOmega_scaling}
		\end{align}
	\end{subequations}
\end{widetext}
These expressions are exact spectral decompositions in which the leading contribution from the coalescing pair has been isolated. The divergence of the QGT components is governed by the scaling of $\mathcal{M}_\mu^{(cm)}$ and $\mathcal{N}_\mu^{(cm)}$, and therefore is not characterised by a single exponent.\\
From Eq.~\eqref{eq:Reg_scaling}, both the rotational  and stretching contribution terms $\mathcal{N}_\mu^{(cm)*}\mathcal{N}_\nu^{(cm)}$ and $\mathcal{M}_\mu^{(cm)*}\mathcal{M}_\nu^{(cm)}$ are generically nonzero at the EP. Their relative scaling determines which contribution dominates the divergence. When the stretching contribution dominates, $\mathrm{Re}\,g_{\mu\nu}$ is driven to $-\infty$, signaling a loss of positive-definiteness and the breakdown of its Riemannian interpretation.\\
The scaling of the geometric quantities near the EP can be summarised in terms of the contributions arising from the coalescing pair. Approaching the EP along a parameter direction $\mu$, with $\epsilon_\mu \to 0$ and $\epsilon_\nu = 0$ for $\nu \neq \mu$, the covariant states scale as
\begin{align}\label{eq:scaling_summary}
    \|\sigma_\mu\|
    &\sim |\epsilon_\mu|^{-(\alpha_\mu+\beta_\mu)}, \\
    \|\Phi_\mu\|
    &\sim |\epsilon_\mu|^{-(\alpha_\mu+\delta_\mu)}.
\end{align}
The corresponding geometric quantities scale as
\begin{align}
    g_{\mu\nu},\;\Omega_{\mu\nu}
    &\sim |\epsilon_\mu|^{-2\bigl(\alpha_\mu+\max(\beta_\mu,\delta_\mu)\bigr)}, \\
    \mathcal{I}_\mu 
    &\sim |\epsilon_\mu|^{-2(\alpha_\mu+\beta_\mu)}.
\end{align}
Thus, the geometric singularity at the EP is characterised by scaling exponents associated with distinct geometric sectors, determined by the structure of the coalescing pair.
%
\section{Application to a Two-Level Non-Hermitian Model}
\label{sec:example}
%
To illustrate the above geometric framework, we apply it to a family of two-level non-Hermitian Hamiltonians that encompasses several physically important models as special cases. In this model, all geometric quantities are obtained in closed analytical form.

\subsection{Model and Quasi-Hermitian Structure}
\label{subsec:model}

We consider the Hamiltonian
\begin{equation}\label{eq:H_model}
    \mathcal{H}(k,\lambda)
    = \begin{pmatrix} i\lambda & d_1(k) \\
      d_2(k) & -i\lambda \end{pmatrix},
\end{equation}
where $k \in [0, 2\pi)$, $\lambda \in \mathbb{R}$ is the gain-loss strength, the variables $(k,\lambda)$ are treated as independent parameters spanning a two-dimensional parameter space, with $\mu,\nu \in \{k,\lambda\}$, and $d_1$, $d_2$ are $2\pi$-periodic complex functions satisfying $d_2(k) = d_1^*(k)$.  Writing
\begin{equation}\label{eq:d_decomp}
    d_1(k) = \Delta(k)\,e^{i\theta(k)},
    \qquad \Delta(k) = |d_1(k)| > 0,
\end{equation}
the eigenvalues are
\begin{equation}\label{eq:energy}
    E_{\pm}(k,\lambda) = \pm\sqrt{\Delta(k)^2 - \lambda^2}\equiv\pm E.
\end{equation}
The system is quasi-Hermitian when $\Delta(k) > |\lambda|$ for all $k$. Let EPs occur at $(k_*, \lambda_*)$ satisfying $\Delta(k_*) = |\lambda_*|$. In the polar gauge, where the Dyson map is chosen as $\eta = G^{1/2}$ with $G = \sum_i \ket{L_i}\bra{L_i}$ [Eq.~\eqref{eq:G_lambda}], we obtain for this system
\begin{equation}\label{eq:eta_model}
	\eta = \begin{pmatrix} a & ib\,e^{i\theta} \\
		-ib\,e^{-i\theta} & a \end{pmatrix},
\end{equation}
where $a$ and $b$ are real functions of $\Delta(k)$ and $\lambda$. The condition $\eta^2 = G$ together with the Hermiticity of $\eta$ gives the relations
\begin{equation}\label{eq:ab}
	a^2 - b^2 = 1, \qquad 2ab = -\frac{\lambda}{E},
	\qquad a^2 + b^2 = \frac{\Delta}{E}.
\end{equation}
The equivalent Hermitian Hamiltonian is
\begin{equation}\label{eq:h_model}
    H = \eta \mathcal{H} \eta^{-1}
      = \begin{pmatrix} 0 & E\,e^{i\theta} \\
        E\,e^{-i\theta} & 0 \end{pmatrix},
\end{equation}
with orthonormal eigenstates
\begin{equation}\label{eq:eigenstates}
    \ket{\Psi_H^{(1)}}
    = \frac{1}{\sqrt{2}}
      \begin{pmatrix} e^{i\theta} \\ 1 \end{pmatrix},
    \qquad
    \ket{\Psi_H^{(2)}}
    = \frac{1}{\sqrt{2}}
      \begin{pmatrix} e^{i\theta} \\ -1 \end{pmatrix},
\end{equation}
corresponding to eigenvalues $\pm E$. Because the eigenstates depend on $k$ solely through the phase $\theta(k)$, we have $\partial_\lambda \ket{\Psi_H^{(m)}} = 0$ for $m = 1, 2$. Consequently, all geometric contributions arising from eigenstate variation (such as the Berry connection) occur only along the $k$-direction. The metric operator $G = \eta^2$ has eigenvalues $g_\pm = \sqrt{(\Delta \pm \lambda)/(\Delta \mp \lambda)}$. For $\lambda > 0$, the eigenvalue $g_-$ vanishes as $\Delta \to \lambda$. This collapse drives the geometric singularity at the EP. In the following subsections, we focus on the target eigenstate $\ket{\Psi_H^{(1)}}$ corresponding to the eigenvalue $+E$, and all geometric quantities are defined with respect to this state unless stated otherwise.

\subsection{Dyson Connection and Covariant States}
\label{subsec:connection_states}
%
To compute the Dyson connection $\Gamma_\mu$, we first determine the parameter derivatives of the matrix elements $a$ and $b$. Using Eq.~\eqref{eq:ab}, we obtain:
\begin{align}\label{eq:ab_derivs}
    \partial_k a &= \frac{\lambda b \Delta'}{2E^2}, \quad
    \partial_k b = \frac{\lambda a \Delta'}{2E^2}, \notag \\
    \partial_\lambda a &= \frac{-\Delta\,b}{2E^2}, \quad
    \partial_\lambda b = \frac{-\Delta\,a}{2E^2}.
\end{align}
where the prime denotes differentiation with respect to $k$, i.e., $\Delta'(k) \equiv d\Delta/dk$. The $k$-derivative of the Dyson map $\eta$ is
\begin{equation}\label{eq:dketa}
    \partial_k \eta = \frac{\lambda \Delta'}{2E^2}
    \begin{pmatrix} b & ia\,e^{i\theta} \\ -ia\,e^{-i\theta} & b \end{pmatrix}
    - \theta'
    \begin{pmatrix} 0 & b\,e^{i\theta} \\ b\,e^{-i\theta} & 0 \end{pmatrix},
\end{equation}
and, the $\lambda$-derivative of $\eta$ is
\begin{equation}\label{eq:dlambdaeta}
    \partial_\lambda \eta = \frac{-\Delta}{2E^2}
    \begin{pmatrix} b & ia\,e^{i\theta} \\ -ia\,e^{-i\theta} & b \end{pmatrix}.
\end{equation}
Using Eqs.~\eqref{eq:dketa} and \eqref{eq:dlambdaeta} together with the identities in Eq.~\eqref{eq:ab}, the Dyson connections [Eq.~\eqref{eq:dyson_connection}] take the form
\begin{align}
	\Gamma_k &= \frac{\lambda \Delta'}{2E^2}\,\tilde{M}
	+ \frac{\lambda\theta'}{2E}\,\tilde{\sigma}
	- ib^2\theta'\,\sigma_z,
	\label{eq:Gamma_k} \\[4pt]
	\Gamma_\lambda &= \frac{-\Delta}{2E^2}\,\tilde{M},
	\label{eq:Gamma_lambda}
\end{align}
where we have introduced the two Hermitian matrices
\begin{equation}\label{eq:Mtilde_def}
	\tilde{M} = \begin{pmatrix} 0 & ie^{i\theta} \\
		-ie^{-i\theta} & 0 \end{pmatrix}, \qquad
	\tilde{\sigma} = \begin{pmatrix} 0 & e^{i\theta} \\
		e^{-i\theta} & 0 \end{pmatrix},
\end{equation}
which satisfy $\tilde{M}^2 = \tilde{\sigma}^2 = \mathbb{I}$ and the anticommutation relation $\{\tilde{M}, \tilde{\sigma}\} = 0$. Separating these into the stretching ($S_\mu$) and rotation ($K_\mu$) components [Eqs.~\eqref{eq:S_def} and \eqref{eq:K_def}] yields:
\begin{align}
    S_k &= \frac{\lambda \Delta'}{2E^2}\,\tilde{M}
         + \frac{\lambda\theta'}{2E}\,\tilde{\sigma},
    \qquad
    K_k = -ib^2\theta'\,\sigma_z,
    \label{eq:SK_k} \\[4pt]
    S_\lambda &= \frac{-\Delta}{2E^2}\,\tilde{M},
    \qquad\qquad\qquad\quad
    K_\lambda = 0.
    \label{eq:SK_lambda}
\end{align}
The stretching component $S_k$ contains contributions proportional to $\Delta'$ and $\theta'$. The rotation component $K_k$ is nonzero only when $\theta' \neq 0$. The metric deformation operator $C_\mu$ [Eq.~\eqref{eq:metric_connection}] takes the form
\begin{align}
    C_k &= \frac{\lambda \Delta'}{2E^2} \tilde{M} 
    + \frac{i\lambda^2 \theta'}{2E^2} \sigma_z 
    + \frac{\lambda \Delta \theta'}{2E^2} \tilde{\sigma}, \\
    C_\lambda &= \frac{-\Delta}{2E^2}\,\tilde{M}.
\end{align}
For the target eigenstate $\ket{\Psi_H^{(1)}}$, the projections and orthogonal fluctuation states defined in Sec.~\ref{sec:decomposition} can be evaluated explicitly. The stretching sector yields:
\begin{subequations}\label{eq:stretching_components}
\begin{align}
    \mathcal{A}_k^{(S)} &= \frac{\lambda\theta'}{2E}, &
    \mathcal{A}_\lambda^{(S)} &= 0, \label{eq:stretching_A} \\
    \ket{\sigma_k} &= \frac{i\lambda \Delta'}{2E^2}\ket{\Psi_H^{(2)}}, &
    \ket{\sigma_\lambda} &= \frac{-i\Delta}{2E^2}\ket{\Psi_H^{(2)}}. \label{eq:stretching_ket}
\end{align}
\end{subequations}
The stretching scalar $\mathcal{I}_\mu$ [Eq.~\eqref{eq:stretching_scalar}] and the system-level scalar evaluate to
\begin{subequations}\label{eq:stretching_invariants}
\begin{align}
    \mathcal{I}_k &= \frac{\lambda^2\theta'^2}{4E^2} + \frac{\lambda^2 \Delta'^2}{4E^4}, \quad
    \mathcal{I}_\lambda = \frac{\Delta^2}{4E^4}, \label{eq:I_k_lambda} \\
    \mathrm{tr}(S_k S_\lambda) &= \mathrm{tr}(C_k C_\lambda) = -\frac{\lambda \Delta \Delta'}{2 E^4}. \label{eq:tr_SC}
\end{align}
\end{subequations}
The rotation projection [Eq.~\eqref{eq:A_S}] vanishes because $\sigma_z$ acts purely off-diagonally on the eigenstates, while the rotation fluctuation state [Eq.~\eqref{eq:kappa}] is nonzero only along $k$,
\begin{subequations}\label{eq:rotation_components}
	\begin{align}
		\mathcal{A}_k^{(K)} &= 0, &
		\mathcal{A}_\lambda^{(K)} &= 0, \label{eq:rotation_A} \\
		\ket{\kappa_k} &= -ib^2\theta'\ket{\Psi_H^{(2)}}, &
		\ket{\kappa_\lambda} &= 0. \label{eq:rotation_ket}
	\end{align}
\end{subequations}
The standard Berry connection [Eq.~\eqref{eq:berry_connection}] and
the orthogonal state $\ket{\phi_\mu}$
[Eq.~\eqref{eq:orthostate}] evaluate to
\begin{subequations}\label{eq:berry_components}
	\begin{align}
		\mathcal{A}_k^{(B)} &= -\frac{\theta'}{2}, &
		\mathcal{A}_\lambda^{(B)} &= 0, \label{eq:berry_A} \\
		\ket{\phi_k} &= \frac{i\theta'}{2}\ket{\Psi_H^{(2)}}, &
		\ket{\phi_\lambda} &= 0. \label{eq:berry_ket}
	\end{align}
\end{subequations}
Therefore, the dressed derivative $\ket{\Phi_\mu}$ [Eq.~\eqref{eq:Phi}] takes the form,
\begin{equation}\label{eq:Phi_eg}
	\ket{\Phi_k} = \frac{i\Delta\theta'}{2E}
	\ket{\Psi_H^{(2)}}, \qquad
	\ket{\Phi_\lambda} = 0.
\end{equation}
\subsection{Non-Hermitian Berry Phase and QGT}
\label{subsec:berry_phase_qgt}

The non-Hermitian Berry connection $\mathcal{A}_\mu$ which governs the adiabatic geometric phase takes the form:
\begin{equation}\label{eq:ANH}
    \mathcal{A}_k
    = -\frac{\theta'}{2}
      \left(1 + \frac{i\lambda}{E}\right), \qquad
    \mathcal{A}_\lambda = 0.
\end{equation}
Integrating over the Brillouin zone $k \in [0, 2\pi)$ gives the complex Berry phase
\begin{equation}\label{eq:gamma_NH_model}
    \gamma
    =-\frac{1}{2}\int_0^{2\pi}\theta'\!\left(1
    +\frac{i\lambda}{E(k)}\right)dk.
\end{equation}
The real part separates as
\begin{equation}\label{eq:gamma_real}
    \Repart\,\gamma
    = -\frac{1}{2}\int_0^{2\pi}\theta'\,dk
    = -\frac{1}{2}\bigl[\theta(2\pi)-\theta(0)\bigr]
    = -\pi\,w,
\end{equation}
where $w = [\theta(2\pi)-\theta(0)]/(2\pi) \in \mathbb{Z}$ is the winding number of the phase $\theta(k)$. It is independent of $\lambda$ and takes integer values.
The imaginary part is
\begin{equation}\label{eq:gamma_imag}
    \Impart\,\gamma
    = -\frac{\lambda}{2}\int_0^{2\pi}
    \frac{\theta'(k)}{E(k)}\,dk,
\end{equation}
which is generally non-quantised and depends on $\lambda$ through $E(k)$. We now examine the geometric structure encoded in the non-Hermitian QGT. Because both covariant states $\ket{\Phi_\mu}$ and $\ket{\sigma_\mu}$ are proportional to $\ket{\Psi_H^{(2)}}$ with purely imaginary coefficients, their inner products are real. Consequently, the QMT $g_{\mu\nu}$ is purely real, and the real part of the non-Hermitian Berry curvature vanishes. The components of QMT [Eq.~\eqref{eq:qgt_g}] are
\begin{align}
    g_{kk} &= \frac{\Delta^2\theta'^2}{4E^2}
             - \frac{\lambda^2 \Delta'^2}{4E^4},
    \label{eq:gkk} \\[4pt]
    g_{\lambda\lambda} &= -\frac{\Delta^2}{4E^4},
    \label{eq:gll} \\[4pt]
    g_{k\lambda} &= \frac{\lambda \Delta \Delta'}{4E^4}
    = g_{\lambda k}.
    \label{eq:gkl}
\end{align}
The metric is generically indefinite. The corresponding line element is real (since $\mathrm{Im}\,g_{\mu\nu}=0$) and can be written as
\begin{equation}
    ds^2 =
    \frac{\Delta^2\theta'^2}{4E^2}\,dk^2
    - \frac{(\lambda \Delta' dk - \Delta d\lambda)^2}{4E^4},
\end{equation}
which makes the indefinite nature of the geometry explicit. The condition $ds^2=0$ defines a critical surface in the $(k,\lambda)$ parameter space. This yields
\begin{equation}
    \Delta^2 \theta'^2 E^2\, dk^2
    =
    (\lambda \Delta' dk - \Delta d\lambda)^2,
\end{equation}
and hence
\begin{equation}
    \frac{d\lambda}{dk}
    =
    \frac{\lambda \Delta'}{\Delta}
    \pm \theta' E,
\end{equation}
define the null directions along which the contributions from the rotation and stretching sectors exactly balance.  Thus, the non-zero component of the non-Hermitian Berry curvature [Eq.~\eqref{eq:qgt_omega}] is
\begin{equation}\label{eq:nh-curve}
    \Omega_{k\lambda}
    = \frac{i\Delta^2\theta'}{2E^3},
\end{equation}
and vanishes when $\theta'=0$. The absence of a real part confirms that the curvature is purely non-Hermitian in origin, arising entirely from the metric deformation encoded in the stretching sector.
%
\subsection{Field Strengths and Curvature Structure}
\label{subsec:field_strengths}
%
We now evaluate the curvatures associated with $\mathcal{A}_\mu^{(B)}$, $\mathcal{A}_\mu^{(K)}$, and $\mathcal{A}_\mu^{(S)}$ [Eqs.~\eqref{eq:omega}--\eqref{eq:FS}]. Since $\mathcal{A}_k^{(B)} = -\theta'/2$ is independent of $\lambda$ and $\mathcal{A}_\lambda^{(B)} = 0$, the standard Berry curvature [Eq.~\eqref{eq:omega}] vanishes identically,
\begin{equation}
	\Omega_{k\lambda}^{(B)} = 0.
\end{equation}
Similarly, the absence of rotation projections, $\mathcal{A}_k^{(K)}=\mathcal{A}_\lambda^{(K)}=0$, implies that the rotation curvature [Eq.~\eqref{eq:FK}] also vanishes,
\begin{equation}\label{eq:f_k}
    \Omega_{k\lambda}^{(K)} = 0.
\end{equation}
In contrast, the stretching sector [Eq.~\eqref{eq:FS}] produces a nontrivial curvature. Using $\mathcal{A}_k^{(S)}=\lambda\theta'/(2E)$ and $\mathcal{A}_\lambda^{(S)}=0$, one finds
\begin{equation}\label{eq:ss_connection}
    \Omega_{k\lambda}^{(S)}
    = -\partial_\lambda\!\left(\frac{\lambda\theta'}{2E}\right)
    = -\frac{\Delta^2\theta'}{2E^3}.
\end{equation}
This result is consistent with the non-Hermitian Berry curvature obtained from the QGT [Eq.~\eqref{eq:nh-curve}], whose real part vanishes while the imaginary part satisfies $\mathrm{Im}\,\Omega_{k\lambda} = -\Omega_{k\lambda}^{(S)}$. Since Eq.~\eqref{eq:nh-curve} is computed from the covariant states $\ket{\Phi_\mu}$ and $\ket{\sigma_\mu}$, while $\Omega_{k\lambda}^{(S)}$ [Eq.~\eqref{eq:ss_connection}] follows from the curl of the connection, their agreement provides an explicit verification of the gauge-invariant identity~\eqref{eq:identity_Im}.\\
At the operator level, the geometric structure reflects the non-commutativity of the stretching components. The corresponding field strength [Eq.~\eqref{eq:FK_result}] evaluates to
\begin{equation}
    \mathcal{F}_{k\lambda}^{(K)}
    = \frac{i\lambda \Delta\theta'}{2E^3}\,\sigma_z.
\end{equation}
This operator-valued curvature is nonzero whenever $\theta$ is $k$--dependent, indicating that successive deformation acting along $k$ and $\lambda$ do not commute. However, its expectation value in the target eigenstate, defined in Eq.~\eqref{eq:FK_onshell_projection}, vanishes because $\sigma_z$ acts purely off-diagonally, consistent with the absence of rotation curvature [Eq.~\eqref{eq:f_k}]. The stretching field strength, evaluated from its definition [Eq.~\eqref{eq:S_field_strength}], vanishes, consistent with Eq.~\eqref{eq:FS_result} of the general framework. These results confirm that the general relations Eqs.~\eqref{eq:FK_result} and~\eqref{eq:FS_result} hold explicitly in this system, with $\mathcal{F}_{k\lambda}^{(K)}$ the only nonzero field strength, arising from the non-commutativity of the stretching components.
\subsection{Scaling Analysis Near the Exceptional Point}
\label{subsec:scaling_ep}
%
Having established the geometric structure through the QGT and the associated field strengths for this system, we now analyse the behaviour of these quantities in the vicinity of the EP. The EP is characterised by the closure of the energy gap defined in Eq.~\eqref{eq:energy}, which vanishes at $(k_*,\lambda_*)$ satisfying $\Delta(k_*) = |\lambda_*|$. To study the behaviour near the EP, we introduce parameter displacements
\begin{equation}
    \epsilon_k \equiv k - k_*, \qquad \epsilon_\lambda \equiv \lambda - \lambda_*,
\end{equation}
and analyse the scaling behaviour along the independent parameter directions.\\
We first consider the scaling of the energy gap. Approaching the EP along the $\lambda$-direction in the quasi-Hermitian region ($\lambda < \lambda_*$), with $\epsilon_\lambda \to 0$ at fixed $k = k_*$, we write $\lambda = \lambda_* - |\epsilon_\lambda|$, which yields
\begin{equation}
    E = \sqrt{\lambda_*^2 - \lambda^2}
      \approx \sqrt{2\lambda_*|\epsilon_\lambda|}
      \;\sim\; |\epsilon_\lambda|^{1/2}.
\end{equation}
Similarly, approaching along the $k$-direction with $\lambda=\lambda_*$ fixed and assuming $\Delta'(k_*) \neq 0$, a Taylor expansion of $\Delta(k)$ gives
\begin{equation}
    E \approx \sqrt{2\lambda_* |\Delta'(k_*)||\epsilon_k|}
    \;\sim\; |\epsilon_k|^{1/2}.
\end{equation}
Thus, the energy gap vanishes with exponent $\alpha_\mu = 1/2$ along both directions, reflecting the square-root closure characteristic of a second-order EP~\cite{kartik2025scaling, heiss2012physics, berry2004physics}. We next examine $\mathcal{M}_\mu^{(nm)}$ and $\mathcal{N}_\mu^{(nm)}$ defined in Eqs.~\eqref{eq:Smu_offdiag} and \eqref{eq:Phi_matrix_element}. For the target eigenstate $|\Psi_H^{(1)}\rangle$, only the off-diagonal elements $\mathcal{M}_\mu^{(21)}$ and $\mathcal{N}_\mu^{(21)}$ appear, corresponding to transitions into the orthogonal subspace. Using their exact forms, one finds
\begin{equation}
    \mathcal{M}_\lambda^{(21)} = \frac{-i\Delta}{E}, \qquad
    \mathcal{M}_k^{(21)} = \frac{i\lambda \Delta'}{E},
\end{equation}
and
\begin{equation}
    \mathcal{N}_\lambda^{(21)} = 0, \qquad
    \mathcal{N}_k^{(21)} = i\Delta\theta'.
\end{equation}
Near the EP,  $\mathcal{M}_\mu^{(21)}$ scale as,
\begin{equation}
    |\mathcal{M}_\mu^{(21)}| \sim E^{-1}
    \sim |\epsilon_\mu|^{-1/2},
\end{equation}
whereas $\mathcal{N}_\mu^{(21)}$ remains finite, showing that the divergence is controlled by the stretching sector. This yields the exponent $\beta_\mu = 1/2$ for both parameter directions. Consequently, the sum of exponents satisfies
\begin{equation}
	\alpha_\mu + \beta_\mu = 1.
\end{equation}
Near the EP, the norm of stretching fluctuation state diverges more rapidly than the norm of dressed derivative state, with
\begin{equation}
    \|\sigma_\mu\| \sim E^{-2} \sim |\epsilon_\mu|^{-1}, 
    \qquad
    \|\Phi_k\| \sim E^{-1} \sim |\epsilon_\mu|^{-1/2}.
\end{equation}
As a result, quantities built from the stretching fluctuation states dominate the geometric response. In particular, the stretching scalar and  system-level scalar scale as
\begin{equation}
    \mathcal{I}_\mu \sim E^{-4} \sim |\epsilon_\mu|^{-2}, \qquad
    \mathrm{tr}(S_k S_\lambda) \sim |\epsilon_\mu|^{-2}
\end{equation}
and the QMT, which is governed by the inner products $\langle \sigma_\mu | \sigma_\nu \rangle$, exhibits the same leading divergence,
\begin{equation}
    g_{\mu\nu} \sim E^{-4} \sim |\epsilon_\mu|^{-2}.
\end{equation}
In contrast, the non-Hermitian Berry curvature involves a mixed contribution between the dressed derivative and the stretching fluctuation states, yielding
\begin{equation}
    \Omega_{k\lambda}
    = \frac{i\Delta^2\theta'}{2E^3}
    \sim E^{-3} \sim |\epsilon_\mu|^{-3/2},
\end{equation}
and therefore diverges more weakly than the QMT. This establishes a clear hierarchy in the geometric response near the EP, with the QMT exhibiting the leading divergence, while the Berry curvature shows a subleading divergence.
%
\subsection{Physical Examples}
\label{subsec:explicit_realisations}
%
The general model encompasses two well-known physical limits, which we now analyse to demonstrate how the geometric quantities reduce in specific cases.
\subsubsection{Non-Hermitian SSH Model}
\label{subsubsec:ssh_model}
\begin{figure}[htbp]
	\centering
	\includegraphics[width=0.49\linewidth]{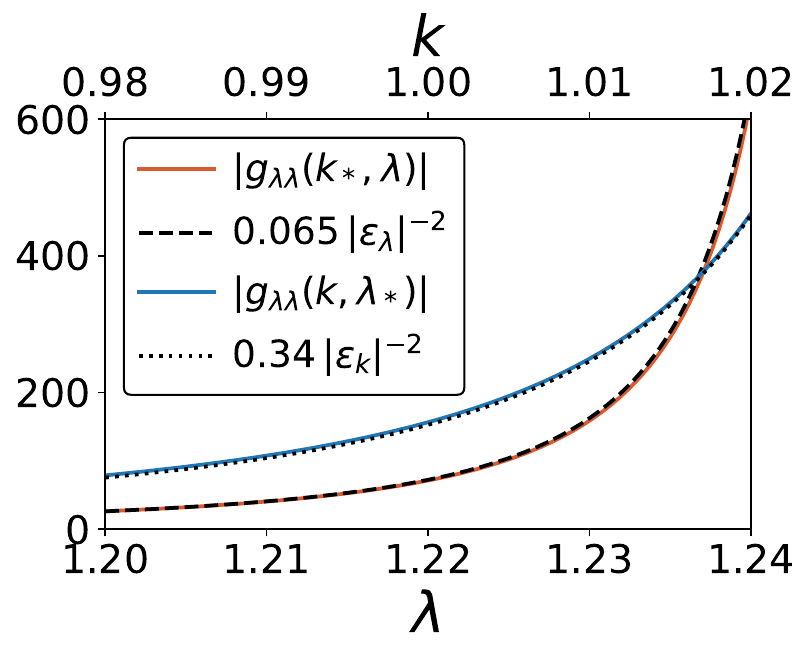}
	\includegraphics[width=0.49\linewidth]{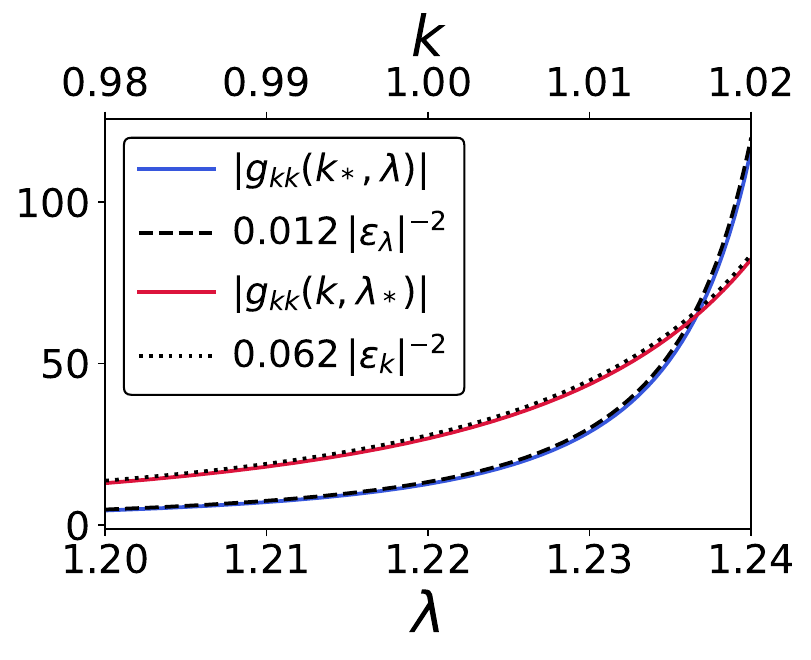}\\
	\includegraphics[width=0.49\linewidth]{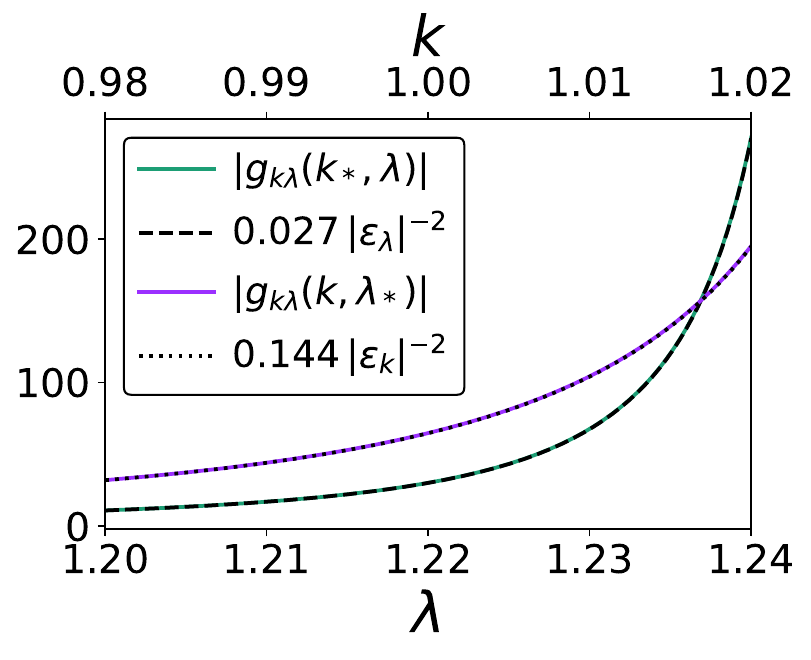}
	\includegraphics[width=0.49\linewidth]{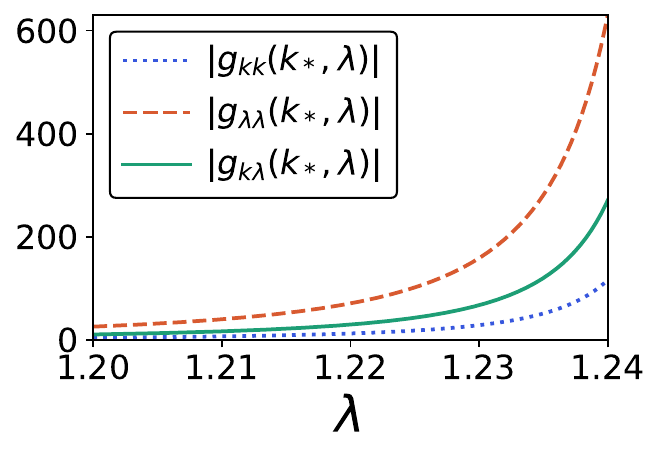}
	\caption{Scaling behaviour and divergence hierarchy of geometric quantities in the SSH model ($t_0 = 1$, $t_1 = 0.5$) as the system approaches the EP at $(k_*, \lambda_*) = \left(\frac{\pi}{3}, 1.25\right)$. The top-left, top-right, and bottom-left panels display the absolute values of the QMT components $|g_{\lambda\lambda}|$, $|g_{kk}|$, and the off-diagonal component $|g_{k\lambda}|$, respectively, along two independent paths to the EP: along $\lambda$ at fixed $k = k_*$ (bottom $\lambda$-axis) and along $k$ at fixed $\lambda = \lambda_*$ (top $k$-axis). Solid lines represent the exact analytical expressions. The dashed and dotted lines correspond to the leading-order asymptotic fits with respect to $\varepsilon_\lambda = \lambda - \lambda_*$ and $\varepsilon_k = k - k_*$, respectively. The numerical fits perfectly confirm the analytically predicted critical exponents: $|g_{\lambda\lambda}|, |g_{kk}|, |g_{k\lambda}| \sim |\varepsilon_\mu|^{-2}$ along both parameter directions, consistent with the scaling law established in Sec.~\ref{subsec:scaling_ep}. The bottom-right panel superimposes all three metric components along the $\lambda$-direction at $k = k_*$, explicitly illustrating the divergence hierarchy $|g_{\lambda\lambda}| > |g_{k\lambda}| > |g_{kk}|$ near the singularity.}
	\label{fig:ssh_geometry}
\end{figure}

%
We first consider a non-Hermitian SSH model~\cite{ lieu2018topological, zeuner2015observation, longhi2023complex},
\begin{equation}
	 \mathcal{H}(k,\lambda)
	= \begin{pmatrix} i\lambda &t_0 + t_1 \cos k \\
		t_0 + t_1 \cos k & -i\lambda \end{pmatrix},
\end{equation}
which is a special case of general 2-level model given in Eq.~\eqref{eq:H_model}. This model describes a one-dimensional lattice with alternating hopping amplitudes and on-site gain/loss. In our parameterisation, this model corresponds to setting the phase to zero, $\theta(k) = 0$, and defining the amplitude as
\begin{equation}
    \Delta(k) = t_0 + t_1 \cos k,
\end{equation}
where $t_0$ and $t_1$ are real constants representing the intra-cell and inter-cell hopping strengths. The energy spectrum is given by
\begin{equation}
    E(k,\lambda) = \sqrt{\Delta^2 - \lambda^2}.
\end{equation}
Because the phase is constant ($\theta' = 0$), the rotation sector vanishes identically. Using the general expressions from Eqs.~\eqref{eq:I_k_lambda} and \eqref{eq:tr_SC}, the stretching scalars and system-level scalars evaluate to:
\begin{align}
    \mathcal{I}_k & = \frac{\lambda^2 t_1^2 \sin^2 k}{4E^4}, \\[4pt]
    \mathcal{I}_\lambda &= \frac{\Delta^2}{4E^4}, \\[4pt]
    \mathrm{tr}(S_k S_\lambda) &= \frac{\lambda \, \Delta\, t_1 \sin k}{2E^4}.
\end{align} 
The non-zero mixed trace $\mathrm{tr}(S_k S_\lambda)$ indicates that metric deformations along the $k$ and $\lambda$ directions are coupled at the operator level. However, the standard Berry connection is zero, and the non-Hermitian Berry curvature [Eq.~\eqref{eq:nh-curve}] vanishes completely:
\begin{equation}
    \Omega_{k\lambda} = 0.
\end{equation}
The non-Hermitian geometry of the system is therefore entirely governed by the stretching sector. Using Eqs.~\eqref{eq:gkk}--\eqref{eq:gkl}, the components of the QMT evaluate to:
\begin{align}
    g_{kk} &= -\frac{\lambda^2 t_1^2 \sin^2 k}{4E^4},  \\
    g_{\lambda\lambda} &= -\frac{\Delta^2}{4E^4},  \\
    g_{k\lambda} &= -\frac{\lambda\, \Delta\, t_1 \sin k}{4E^4}.
\end{align}
Notably, both $g_{kk} \le 0$ and $g_{\lambda\lambda} \le 0$, so the QMT is negative semi-definite. The absence of a rotation component means that there are no balancing null directions, and the geometry reflects pure metric deformation without any associated curvature.\\
Near an EP, at $(k_*, \lambda_*)$ satisfying $\Delta(k_*) = |\lambda_*|$, the energy gap closes as $E \sim |\epsilon_\mu|^{1/2}$ along both parameter directions (assuming $\Delta'(k_*) \neq 0$). Consequently, all components of the QMT exhibit the leading divergence
\begin{equation}
g_{kk},\; g_{\lambda\lambda},\; g_{k\lambda} \;\sim\; |\epsilon_\mu|^{-2},
\end{equation}
consistent with the general scaling law established in Sec.~\ref{subsec:scaling_ep}. The non-Hermitian Berry curvature vanishes identically throughout the parameter space, and the SSH model therefore provides a minimal example in which the geometric singularity at the EP is entirely encoded in QMT. These scaling behaviours and the divergence hierarchy are illustrated in Fig.~\ref{fig:ssh_geometry}.
\subsubsection{HN Model}
\label{subsubsec:hn_model}
%
The celebrated HN model~\cite{hatano1996localization, longhi2023complex}, described by
\begin{equation}
	\mathcal{H}(k,\lambda)
	= \begin{pmatrix} i\lambda & R_0\,e^{ik} \\
	R_0\,e^{-ik} & -i\lambda \end{pmatrix},
\end{equation}
is also a special case of the model in Eq.~\eqref{eq:H_model}, with
\begin{equation}
	\Delta(k) = R_0, \qquad \theta(k) = k.
\end{equation}
Because $\Delta' = 0$, the energy spectrum is completely independent of $k$, yielding flat real energy bands as long as the system remains in the quasi-Hermitian regime ($R_0 > |\lambda|$):
\begin{equation}
    E(\lambda) = \sqrt{R_0^2 - \lambda^2}.
\end{equation}
A distinctive consequence of $\Delta' = 0$ is that the EP condition $\Delta(k_*) = |\lambda_*|$ reduces to $R_0 = |\lambda_*|$, which is independent of $k$. The set of EPs is therefore an entire line in parameter space, $\{(k, R_0) : k \in [0, 2\pi)\}$. Consequently, the only meaningful displacement from the EP is $\epsilon_\lambda = \lambda - R_0$, and no analogue of $\epsilon_k$ exists.\\
The stretching scalars $\mathcal{I}_\mu$ and the system-level scalars for this model simplify significantly. Because the amplitude is constant ($\Delta'=0$), the mixed trace vanishes, yielding:
\begin{align}
    \mathcal{I}_k &=  \frac{\lambda^2}{4E^2}, \\[4pt]
    \mathcal{I}_\lambda &=  \frac{R_0^2}{4E^4}, \\[4pt]
    \mathrm{tr}(S_k S_\lambda) &= 0.
\end{align}
The vanishing mixed trace $\mathrm{tr}(S_k S_\lambda) = 0$ indicates that metric deformations along $k$ and $\lambda$ are completely decoupled at the operator level. Correspondingly, the off-diagonal QMT component $g_{k\lambda}$ vanishes, and the QMT is strictly diagonal and indefinite:
\begin{equation}
    g_{kk} = \frac{R_0^2}{4E^2}, \qquad
    g_{\lambda\lambda} = -\frac{R_0^2}{4E^4}.
\end{equation}
\begin{figure}[htbp]
	\centering
	\includegraphics[width=0.49\linewidth]{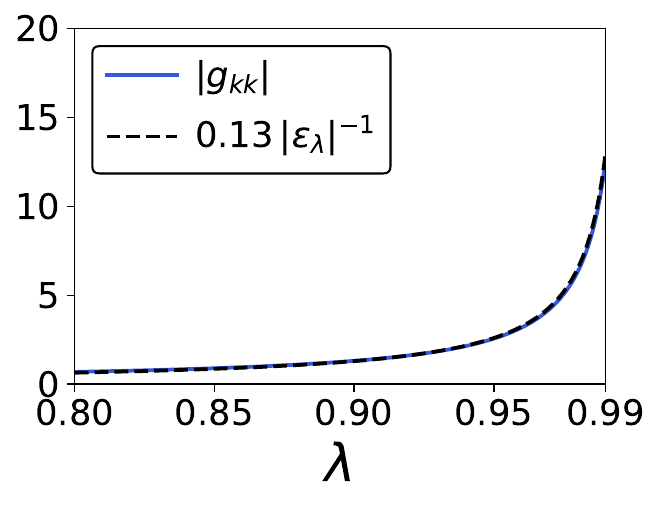}
	\includegraphics[width=0.49\linewidth]{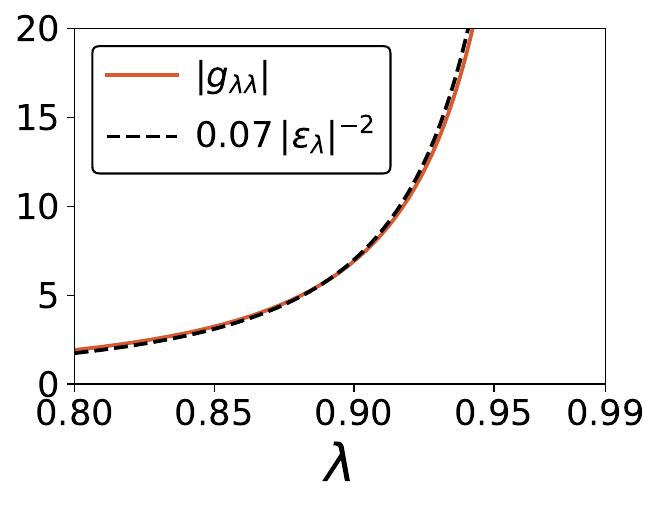}\\
	\includegraphics[width=0.49\linewidth]{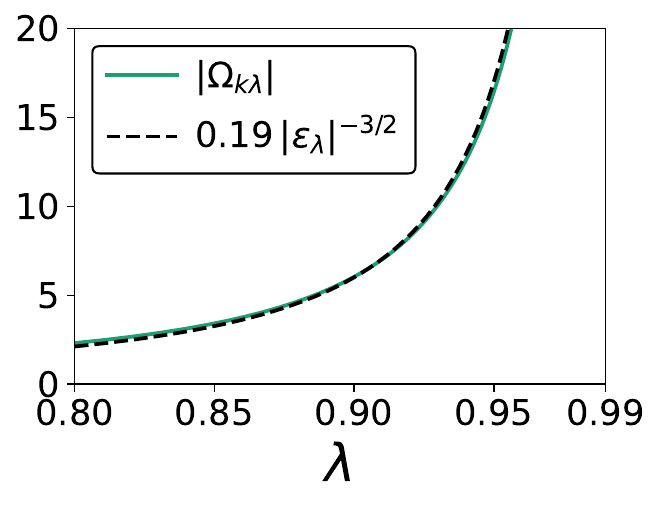}
	\includegraphics[width=0.49\linewidth]{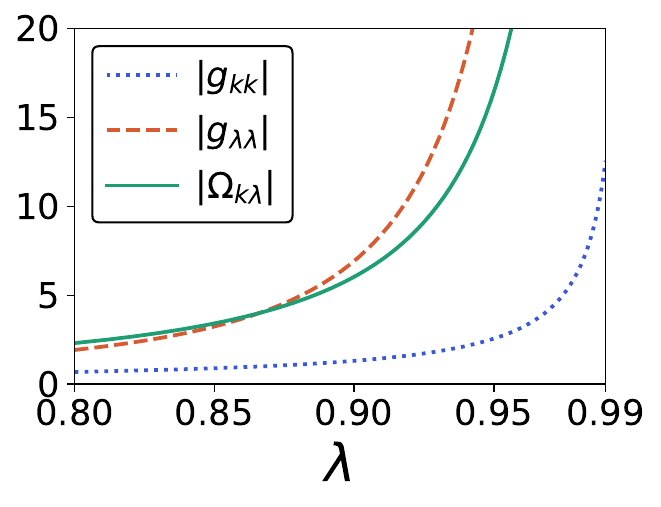}
	\caption{Scaling behaviour and divergence hierarchy of geometric quantities in the HN model ($R_0 = 1$) as the system approaches the EP at $\lambda \to 1$. The top-left, top-right, and bottom-left panels display the absolute values of the QMT components $|g_{kk}|$ and $|g_{\lambda\lambda}|$, and the non-Hermitian Berry curvature $|\Omega_{k\lambda}|$, respectively, as functions of the parameter $\lambda$. Solid coloured lines represent the exact analytical expressions. The black dashed lines correspond to the leading-order asymptotic fits with respect to $\varepsilon_\lambda = \lambda -1$. The numerical fits perfectly confirm the analytically predicted critical exponents: $|g_{kk}| \sim |\varepsilon_\lambda|^{-1}$, $|g_{\lambda\lambda}| \sim |\varepsilon_\lambda|^{-2}$, and $|\Omega_{k\lambda}| \sim |\varepsilon_\lambda|^{-3/2}$. The bottom-right panel superimposes all three quantities, explicitly illustrating the divergence hierarchy established in Sec.~\ref{subsec:scaling_ep}, where the metric deformation along $\lambda$ direction ($|g_{\lambda\lambda}|$) dominates the geometric response near the singularity.}
	\label{fig:hn_geometry}
\end{figure}
\begin{table*}[t]
	\centering
	\renewcommand{\arraystretch}{1.5}
	\setlength{\tabcolsep}{0pt}
	\begin{ruledtabular}
		\begin{tabular*}{\columnwidth}{@{\extracolsep{\fill}}lccc}
			\textbf{Quantity} &
			\textbf{General model} &
			\textbf{SSH} ($\theta'=0$) &
			\textbf{HN} ($\Delta'=0$) \\
			\hline
			\multicolumn{4}{l}{}\\[0pt]
			Amplitude, phase & $\Delta(k),\ \theta(k)$ & $t_0+t_1\cos k,\ \ 0$ & $R_0,\ \ k$ \\
			Energy & $\sqrt{\Delta^2-\lambda^2}$ & $\sqrt{(t_0+t_1\cos k)^2-\lambda^2}$ & $\sqrt{R_0^2-\lambda^2}$ \\[4pt]
			\hline
			\multicolumn{4}{l}{\textit{Stretching and rotation }}\\[-5pt]
			\multicolumn{4}{l}{\textit{components}}\\[2pt]
			$S_k$ & $\dfrac{\lambda\Delta'}{2E^2}\tilde M+\dfrac{\lambda\theta'}{2E}\tilde\sigma$ & $\dfrac{\lambda\Delta'}{2E^2}\tilde M$ &
			$\dfrac{\lambda}{2E}\tilde\sigma$ \\[8pt]
			$S_\lambda$ & $-\dfrac{\Delta}{2E^2}\tilde M$ & $-\dfrac{\Delta}{2E^2}\tilde M$ & $-\dfrac{R_0}{2E^2}\tilde M$ \\[8pt]
			$K_k$ & $-ib^2\theta'\,\sigma_z$ & $0$ & $-ib^2\,\sigma_z$ \\[4pt]
			\hline
			\multicolumn{4}{l}{\textit{Covariant states}}\\[-5pt]
			$\ket{\sigma_k}$ & $\dfrac{i\lambda\Delta'}{2E^2}\ket{\Psi_H^{(2)}}$ & $\dfrac{i\lambda\Delta'}{2E^2}\ket{\Psi_H^{(2)}}$ & $0$ \\[8pt]
			$\ket{\sigma_\lambda}$ & $-\dfrac{i\Delta}{2E^2}\ket{\Psi_H^{(2)}}$ & $-\dfrac{i\Delta}{2E^2}\ket{\Psi_H^{(2)}}$ & $-\dfrac{iR_0}{2E^2}\ket{\Psi_H^{(2)}}$ \\[8pt]	
			$\ket{\Phi_k}$ & $\dfrac{i\Delta\theta'}{2E}\ket{\Psi_H^{(2)}}$ & $0$ & $\dfrac{iR_0}{2E}\ket{\Psi_H^{(2)}}$ \\[6pt]
			\hline
			\multicolumn{4}{l}{\textit{Projections and Berry}}\\[-5pt]
			\multicolumn{4}{l}{\textit{connection}}\\[2pt]
			$\mathcal{A}_k^{(S)}$ & $\dfrac{\lambda\theta'}{2E}$ & $0$ & $\dfrac{\lambda}{2E}$ \\[6pt]
			$\mathcal{A}_k^{(B)}$ & $-\dfrac{\theta'}{2}$ & $0$ & $-\dfrac12$ \\[6pt]
			\hline
			\multicolumn{4}{l}{\textit{Stretching scalar and}}\\[-5pt]
			\multicolumn{4}{l}{\textit{system-level scalar}}\\[2pt]
			$\mathcal{I}_k$ & $\dfrac{\lambda^2\theta'^2}{4E^2}+\dfrac{\lambda^2\Delta'^2}{4E^4}$ & $\dfrac{\lambda^2 t_1^2\sin^2 k}{4E^4}$ & $\dfrac{\lambda^2}{4E^2}$ \\[8pt]
			$\mathcal{I}_\lambda$ & $\dfrac{\Delta^2}{4E^4}$ & $\dfrac{\Delta^2}{4E^4}$ & $\dfrac{R_0^2}{4E^4}$ \\[8pt]
			$\mathrm{tr}(S_kS_\lambda)$ & $-\dfrac{\lambda\Delta\Delta'}{2E^4}$ & $\dfrac{\lambda\Delta t_1\sin k}{2E^4}$ & $0$ \\[6pt]
			\hline
			\multicolumn{4}{l}{\textit{Components of QMT}}\\[2pt]
			$g_{kk}$ & $\dfrac{\Delta^2\theta'^2}{4E^2}-\dfrac{\lambda^2\Delta'^2}{4E^4}$ & $-\dfrac{\lambda^2 t_1^2\sin^2 k}{4E^4}$ & $\dfrac{R_0^2}{4E^2}$ \\[8pt]
			$g_{\lambda\lambda}$ & $-\dfrac{\Delta^2}{4E^4}$ & $-\dfrac{\Delta^2}{4E^4}$ & $-\dfrac{R_0^2}{4E^4}$ \\[8pt]
			$g_{k\lambda}$ & $\dfrac{\lambda\Delta\Delta'}{4E^4}$ & $-\dfrac{\lambda\Delta t_1\sin k}{4E^4}$ & $0$ \\[6pt]
			Signature & indefinite & negative\ semi-definite & indefinite \\[8pt]
			\hline
			\multicolumn{4}{l}{\textit{Curvature and field strength}}\\[2pt]
			$\Omega_{k\lambda}$ & $\dfrac{i\Delta^2\theta'}{2E^3}$ & $0$ & $\dfrac{iR_0^2}{2E^3}$ \\[8pt]
			$\mathcal{F}_{k\lambda}^{(K)}$ & $\dfrac{i\lambda\Delta\theta'}{2E^3}\sigma_z$ & $0$ & $\dfrac{i\lambda R_0}{2E^3}\sigma_z$ \\[6pt]
		\end{tabular*}
	\end{ruledtabular}
\end{table*}
\begin{table*}[t]
	\centering
	\renewcommand{\arraystretch}{1.5}
	\setlength{\tabcolsep}{0pt}
	\begin{ruledtabular}
		\begin{tabular*}{\columnwidth}{@{\extracolsep{\fill}}lccc}
			\textbf{Quantity} &
			\textbf{General model} &
			\textbf{SSH} ($\theta'=0$) &
			\textbf{HN} ($\Delta'=0$) \\
			\hline
			\multicolumn{4}{l}{\textit{Berry phase, EP and scaling }}\\[-5pt]
			\multicolumn{4}{l}{\textit{behavior}}\\[2pt]
			$\Repart\,\gamma$ & $-\pi w$ & $0$ & $-\pi$ \\
			$\Impart\,\gamma$ & $-\dfrac{\lambda}{2}\displaystyle\int_0^{2\pi}\!\dfrac{\theta'}{E}\,dk$ & $0$ & $-\dfrac{\pi\lambda}{E}$ \\[8pt]
			EP & $\Delta(k_*)=|\lambda_*|$ & $t_0+t_1\cos k_*=|\lambda_*|$ & $R_0=|\lambda_*|$ \\[4pt]
			QMT scaling & $g_{\mu\nu}\sim|\epsilon_\mu|^{-2}$ & $g_{\mu\nu}\sim|\epsilon_\mu|^{-2}$ & $g_{\lambda\lambda}\sim|\epsilon_\lambda|^{-2}$ \\
			Anomalous scaling & --- & --- & $g_{kk}\sim|\epsilon_\lambda|^{-1}$ \\
			Curvature scaling & $\Omega_{k\lambda}\sim|\epsilon_\mu|^{-3/2}$ & --- & $\Omega_{k\lambda}\sim|\epsilon_\lambda|^{-3/2}$ \\
		\end{tabular*}
	\end{ruledtabular}
	\caption{Complete geometric structure of the general two-level non-Hermitian model [Eq.~\eqref{eq:H_model}], with its two physical models: the non-Hermitian SSH model [$\theta=0$, $\Delta = t_0+t_1\cos k$, so $\theta'=0$] and the HN model [$\Delta=R_0$, $\theta=k$, so $\Delta'=0$]. Every entry in the two special cases follows from the general column by setting $\theta'=0$ (SSH) or $\Delta'=0$ (HN). In the SSH model the geometry is entirely metric-deformation driven, since $\ket{\Phi_\mu}=0$, while in the HN model the stretching fluctuation state along $k$ vanishes, $\ket{\sigma_k}=0$. Quantities that vanish identically in the general model are omitted: the rotation projection $\mathcal{A}_\mu^{(K)}=0$, the $\lambda$-components $K_\lambda=\ket{\Phi_\lambda}=\ket{\kappa_\lambda}=0$, and $\mathcal{A}_\lambda^{(S)}=\mathcal{A}_\lambda^{(B)}=0$.}
	\label{tab:model_comparison}
\end{table*}
Here, the QMT exhibits a positive signature along the $k$-direction (driven by phase winding) and a negative signature along the $\lambda$-direction (driven by the metric deformation from the gain-loss variation). This is further reflected in the anisotropic divergence of the QMT components, with $g_{\lambda\lambda} \sim E^{-4}$ and $g_{kk} \sim E^{-2}$. The dominant divergence of $g_{\lambda\lambda}$, which originates purely from the metric deformation ($\ket{\Phi_\lambda} = 0$) and is negative, confirms that the metric deformation governs the leading geometric response near the EP, consistent with the indefinite character of the QMT established in Sec.~\ref{sec:qgt}.\\
Unlike the SSH model, the presence of a non-zero $\theta'$ generates a non-trivial non-Hermitian Berry curvature. Using Eq.~\eqref{eq:nh-curve}, we find:
\begin{equation}
    \Omega_{k\lambda} = \frac{i R_0^2}{2E^{3}}.
\end{equation}
This curvature is purely imaginary, confirming its origin from the metric deformation rather than standard unitary holonomy. Furthermore, integrating the non-Hermitian Berry connection over the Brillouin zone [Eq.~\eqref{eq:gamma_NH_model}] yields the complex Berry phase:
\begin{equation}
    \gamma = -\pi \left( 1 + \frac{i\lambda}{E} \right).
\end{equation}
The real part is strictly quantised to $-\pi$ (corresponding to a winding number $w=1$), while the imaginary part grows non-linearly with the gain-loss strength $\lambda$.\\
As expected from the general scaling analysis in Sec.~\ref{subsec:scaling_ep}, these geometric quantities diverge as the system approaches the EP at $|\lambda| \to R_0$. The QMT component $g_{\lambda\lambda}$ exhibits the strongest divergence ($\sim |\epsilon_\lambda|^{-2}$), while the curvature diverges more weakly ($\sim |\epsilon_\lambda|^{-3/2}$). Notably, the diagonal component $g_{kk}$ exhibits a weaker divergence $\sim |\epsilon_\lambda|^{-1}$ rather than the generic $|\epsilon_\lambda|^{-2}$ scaling predicted in Sec.~\ref{subsec:scaling_ep}. This arises because the constant amplitude $\Delta(k) = R_0$ implies $\Delta'(k) = 0$, so the stretching fluctuation state along the $k$-direction vanishes identically, $\ket{\sigma_k} = 0$. The dominant $|\epsilon_\lambda|^{-2}$ contribution to $g_{kk}$, which in the general case arises from $\|\sigma_k\|^2$, is therefore absent, and $g_{kk}$ is governed entirely by the dressed derivative state, $g_{kk} = \|\Phi_k\|^2 \sim |\epsilon_\lambda|^{-1}$. This hierarchy is illustrated in Fig.~\ref{fig:hn_geometry}. The weaker scaling ($g_{kk} \sim |\epsilon_\lambda|^{-1}$) coincides with the hyperbolic divergence of the QMT measured at a second-order EP in exciton-polariton microcavities~\cite{liao2021experimental}, where the QMT was found to scale as $q^{-1}$ with the parameter displacement $q$ from the EP. The complete geometric structure of the general model, together with its SSH and HN limits, is summarised in Table~\ref{tab:model_comparison}.
%
\section{Conclusion}
\label{sec:conclusion}
%
In this work, we have developed a gauge-covariant geometric framework for non-Hermitian quantum systems in the quasi-Hermitian regime by elevating the Dyson map from a similarity transformation to a central geometric object on parameter space. Its parameter dependence defines the Dyson connection $\Gamma_\mu = (\partial_\mu\eta)\eta^{-1}$, whose decomposition into a Hermitian component (stretching) $S_\mu$ and an anti-Hermitian component (rotation) $K_\mu$ cleanly separates the physical metric deformation of the state space from the unitary redundancy inherent in the choice of the Dyson map. This separation resolves a persistent difficulty of biorthogonal quantum mechanics~\cite{brody2014biorthogonal}: it provides an explicit, basis-independent criterion for distinguishing genuine non-Hermitian geometry from artifacts of the biorthogonal basis choice.\\
Building on this decomposition, we constructed two manifestly gauge-covariant states, the dressed derivative state $\ket{\Phi_\mu}$ and the stretching fluctuation state
$\ket{\sigma_\mu}$, whose inner products generate the complete gauge-invariant geometry of the system. From these we derived the complex non-Hermitian Berry connection and Berry phase~\cite{garrison1988complex, longhi2023complex}, which are gauge-invariant quantities. A key finding is that the imaginary part of the non-Hermitian Berry connection is fixed entirely by the gauge-invariant stretching projection $\mathcal{A}_\mu^{(S)}$ and governs the path-dependent amplification or attenuation of state norms. A nonzero $\mathcal{A}_\mu^{(S)}$ gives the geometric phase complex; the resulting imaginary part has direct physical consequences, controlling, for instance, whether the spectral phase transition of a slowly cycled non-Hermitian system is sharp or imperfect~\cite{longhi2023complex}.\\
We further derived the non-Hermitian QGT and expressed it entirely in terms of the covariant states $\ket{\Phi_\mu}$ and $\ket{\sigma_\mu}$. The resulting QMT is generically indefinite and its imaginary part has no Hermitian counterpart. Although previous works~\cite{chen2024quantum, hu2025quantum} have noted the indefinite QMT and the resulting pseudo-Riemannian geometry by using the left-right formulation of the QGT, our framework supplies their physical origin: the negative contribution to the QMT arises entirely from the stretching fluctuation state $\ket{\sigma_\mu}$. This further yields, in closed form, a critical (null) surface within the quasi-Hermitian region, where $\mathrm{Re}(ds^2)$ changes sign marking the onset of indefinite geometry as a precursor to the exceptional point. The flatness of the Dyson connection further yields the operator-level relations $\mathcal{F}_{\mu\nu}^{(K)} = [S_\mu, S_\nu]$ and $\mathcal{F}_{\mu\nu}^{(S)} = 0$, identifying the non-commutativity of metric deformations along different parameter directions as the algebraic origin of non-Hermitian geometric curvature.\\
Near an exceptional point, the framework traces the geometric singularity to the divergence of the covariant states and yields a hierarchy of scaling laws characterised by independent exponents for the stretching and dressed-derivative sectors. For a general two-level non-Hermitian model, the QMT exhibits the leading divergence $\sim |\epsilon_\mu|^{-2}$, while the non-Hermitian Berry curvature diverges only as $|\epsilon_\mu|^{-3/2}$, establishing a distinct divergence hierarchy near the EP. The non-Hermitian SSH and HN models follow as exact analytical limits in which all geometric quantities are obtained in closed form: the SSH model realises a purely metric-deformation-driven, negative semi-definite geometry with vanishing curvature, while the HN model exhibits an anomalous subleading metric divergence $g_{kk} \sim |\epsilon_\lambda|^{-1}$, coinciding with the hyperbolic divergence of the quantum metric measured at a second-order EP in exciton-polariton microcavities~\cite{liao2021experimental}. While these specific exponents follow from the square-root structure of a second-order EP, the spectral decompositions underlying the scaling analysis hold for EPs of arbitrary order.\\
Several directions follow naturally from this work. The scaling analysis can be extended to higher-order EPs, where multiple partner states coalesce and the interplay of several divergent off-diagonal matrix elements may produce richer exponent structures~\cite{heiss2012physics, bergholtz2021exceptional}. The gauge-covariant states and the stretching scalars introduced here are constructed from measurable quantities, state norms and overlaps, and are therefore natural targets for experimental platforms in which the quantum geometry of non-Hermitian systems is already accessible, including exciton-polariton microcavities~\cite{liao2021experimental, cuerda2024observation} and photonic systems with engineered gain and loss~\cite{ozdemir2019parity, el2018non}. Extending the present construction beyond the quasi-Hermitian regime, where the metric loses positive-definiteness and the Dyson map becomes non-invertible, remains an open problem~\cite{mostafazadeh2010pseudo}. Finally, since the stretching sector supplies gauge-invariant geometric data beyond the non-Hermitian Berry curvature, it would be interesting to examine its role in the topological classification of non-Hermitian bands~\cite{bergholtz2021exceptional, kawabata2019symmetry}, where the interplay between metric deformation and topology remains largely unexplored.
%
\section{Acknowledgment}
%
G.D. acknowledges UGC-JRF, New Delhi, India, for JRF fellowship. S.S. acknowledges NBCFDC, New Delhi, India, for JRF fellowship. BPM acknowledges the PDF research grant for faculty member under the IoE Scheme for the year 2025--26 of Banaras Hindu University.
%
%
\bibliography{ref2}
\appendix
%
\section{Gauge Freedom of the Dyson Map and Gauge Structure}
\label{app:gauge_structure}
%
As introduced in Sec.~\ref{sec:geometric_framework}, the Dyson map $\eta$ is not unique. If $\eta$ is a Dyson map satisfying $H = \eta \mathcal{H} \eta^{-1}$ with $H = H^\dagger$, then for any unitary operator $U(\lambda^\mu)$, the transformed map
\begin{equation}\label{eq:app_dyson_equiv}
    \bar{\eta} = U\eta, \qquad U^\dagger U = \mathbb{I},
\end{equation}
is also a valid Dyson map. Because $U$ acts on the $N$-dimensional Hilbert space, this redundancy mathematically constitutes a local $U(N)$ gauge freedom at the operator level, which reduces to a $U(1)$ gauge freedom when projected onto individual non-degenerate eigenstates. In this appendix, we systematically derive how all geometric quantities transform under the Dyson gauge freedom $\eta \to \bar{\eta}$, classifying them as gauge-invariant, gauge-covariant, and gauge-dependent. 
%
\subsection{Transformations of Operators and States}
%
Under the transformation $\eta \to \bar{\eta}$, the equivalent Hermitian Hamiltonian transforms by unitary conjugation:
\begin{equation}
    \bar{H} = \bar{\eta} \mathcal{H} \bar{\eta}^{-1} = U \eta \mathcal{H} \eta^{-1} U^\dagger = U H U^\dagger.
\end{equation}
Consequently, the orthonormal eigenstates $\ket{\Psi_H}$ of $H$ transform covariantly:
\begin{equation}
    \ket{\bar{\Psi}_H} = U \ket{\Psi_H}.
\end{equation}
The right and left eigenstates of the non-Hermitian Hamiltonian $\mathcal{H}$, however, are completely independent of the Dyson map and remain strictly invariant:
\begin{align}
    \ket{\bar{R}} &= \bar{\eta}^{-1}\ket{\bar{\Psi}_H} = \eta^{-1} U^\dagger U \ket{\Psi_H} = \eta^{-1}\ket{\Psi_H} = \ket{R}, \\
    \bra{\bar{L}} &= \bra{\bar{\Psi}_H}\bar{\eta} = \bra{\Psi_H} U^\dagger U \eta = \bra{\Psi_H}\eta = \bra{L}.
\end{align}
The metric operator $G = \eta^\dagger \eta$ is manifestly gauge-invariant:
\begin{equation}
    \bar{G} = \bar{\eta}^\dagger \bar{\eta} = \eta^\dagger U^\dagger U \eta = \eta^\dagger \eta = G.
\end{equation}
Because $G$ is invariant, the metric deformation operator defined in Eq.~\eqref{eq:metric_connection} is also strictly gauge-invariant:
\begin{equation}
    \bar{C}_\mu = \frac{1}{2} \bar{G}^{-1} \partial_\mu \bar{G} = \frac{1}{2} G^{-1} \partial_\mu G = C_\mu.
\end{equation}
%
\subsection{The Dyson Connection and its Components}
%
The Dyson connection transforms as a standard gauge connection:
\begin{align}
	\bar{\Gamma}_\mu &= (\partial_\mu \bar{\eta})\bar{\eta}^{-1} = \bigl[(\partial_\mu U)\eta + U(\partial_\mu \eta)\bigr]\eta^{-1} U^\dagger \nonumber \\
	&= \omega_\mu + U \Gamma_\mu U^\dagger,
\end{align}
where we define the pure-gauge one-form $\omega_\mu \equiv (\partial_\mu U)U^\dagger$ with $\omega_\mu + \omega_\mu^\dagger = 0$. For the stretching component [Eq.~\eqref{eq:S_def}]:
\begin{equation}
    \bar{S}_\mu = \frac{1}{2}(\bar{\Gamma}_\mu + \bar{\Gamma}_\mu^\dagger) = U S_\mu U^\dagger.
\end{equation}
Thus, $S_\mu$ transforms covariantly as an adjoint field. This is consistent with its relation to the invariant metric deformation operator $C_\mu$, which gives $\bar{S}_\mu = \bar{\eta} C_\mu \bar{\eta}^{-1} = U \eta C_\mu \eta^{-1} U^\dagger = U S_\mu U^\dagger$.\\
For the rotation component [Eq.~\ref{eq:K_def}]:
\begin{equation}
    \bar{K}_\mu = \frac{1}{2}(\bar{\Gamma}_\mu - \bar{\Gamma}_\mu^\dagger) = \omega_\mu + U K_\mu U^\dagger.
\end{equation}
$K_\mu$ has the pure-gauge term, transforming inhomogeneously as a gauge connection. Because $\Gamma_\mu$ and $K_\mu$ both have this pure-gauge shift $\omega_\mu$, their associated covariant derivatives transform covariantly as operators. The full covariant derivative $D_\mu$ [Eq.~\eqref{eq:full_cov}] and the $K$-covariant derivative $D_\mu^{(K)}$ [Eq.~\eqref{eq:K_cov}] transform as:
\begin{align}
    \bar{D}_\mu &= U D_\mu U^\dagger, \\
    \bar{D}_\mu^{(K)} &= U D_\mu^{(K)} U^\dagger.
\end{align}
Therefore, their action on a covariant state yields another covariant state. Similarly, the adjoint $K$-covariant derivative $\mathcal{D}_\mu^{(K)}$ [Sec.~\ref{sec:field_str}], which acts on operators, also transforms covariantly. For any covariant operator $O$ (which transforms as $\bar{O} = U O U^\dagger$), its covariant derivative transforms as:
\begin{equation}\label{eq:dd}
    \bar{\mathcal{D}}_\mu^{(K)}(\bar{O}) = U \bigl( \mathcal{D}_\mu^{(K)} O \bigr) U^\dagger.
\end{equation}
\subsection{Projected States}
%
The orthogonal projector $\mathcal{Q} = \mathbb{I} - \ket{\Psi_H}\bra{\Psi_H}$ transforms covariantly: $\bar{\mathcal{Q}} = U \mathcal{Q} U^\dagger$. The stretching fluctuation state $\ket{\sigma_\mu}$ [Eq.~\eqref{eq:sigma}] transforms covariantly without acquiring any gauge artifact:
\begin{equation}
	\ket{\bar{\sigma}_\mu} = \bar{\mathcal{Q}} \bar{S}_\mu \ket{\bar{\Psi}_H} = U \mathcal{Q} S_\mu \ket{\Psi_H} = U \ket{\sigma_\mu}.
\end{equation}
In contrast, the state derivative $\ket{\phi_\mu}$ [Eq.~\eqref{eq:orthostate}] and the rotation fluctuation state $\ket{\kappa_\mu}$ [Eq.~\eqref{eq:kappa}] both acquire the same gauge artifact:
\begin{align}
	\ket{\bar{\phi}_\mu} &= U \ket{\phi_\mu} + U \ket{\tilde{\omega}_\mu}, \\
	\ket{\bar{\kappa}_\mu} &= U \ket{\kappa_\mu} + U \ket{\tilde{\omega}_\mu},
\end{align}
where $\ket{\tilde{\omega}_\mu} \equiv \mathcal{Q} \tilde{\omega}_\mu \ket{\Psi_H}$ with $\tilde{\omega}_\mu \equiv U^\dagger \omega_\mu U = U^\dagger \partial_\mu U$.\\
The dressed derivative state $\ket{\Phi_\mu} = \ket{\phi_\mu} - \ket{\kappa_\mu}$ subtracts these two quantities, exactly canceling the inhomogeneous gauge artifact. It therefore transforms covariantly:
\begin{equation}\label{eq:app_phi}
    \ket{\bar{\Phi}_\mu} = U \ket{\Phi_\mu}.
\end{equation}
\subsection{Scalar projections, Berry Phase and the system scalars}\label{subsec:connections}
%
The scalar projections defined in Eqs.~\eqref{eq:A_S} and~\eqref{eq:berry_connection} transform as:
\begin{align}
	\bar{\mathcal{A}}_\mu^{(B)} &= i\braket{\bar{\Psi}_H}{\partial_\mu \bar{\Psi}_H} = \mathcal{A}_\mu^{(B)} - a_\mu, \\
	\bar{\mathcal{A}}_\mu^{(K)} &= -i\mel{\bar{\Psi}_H}{\bar{K}_\mu}{\bar{\Psi}_H} = \mathcal{A}_\mu^{(K)} + a_\mu, \\
	\bar{\mathcal{A}}_\mu^{(S)} &= \mel{\bar{\Psi}_H}{\bar{S}_\mu}{\bar{\Psi}_H} = \mathcal{A}_\mu^{(S)},
\end{align}
where real gauge-shift scalar $a_\mu \equiv -i\mel{\Psi_H}{\tilde{\omega}_\mu}{\Psi_H}$. The stretching projection is intrinsically gauge-invariant. The standard Berry connection and the rotation projection are individually gauge-dependent, shifting by equal and opposite amounts. Their sum is therefore gauge-invariant:
\begin{equation}
    \bar{\mathcal{A}}_\mu^{(B)} + \bar{\mathcal{A}}_\mu^{(K)} = \mathcal{A}_\mu^{(B)} + \mathcal{A}_\mu^{(K)}.
\end{equation}
Consequently, the full non-Hermitian Berry connection $\mathcal{A}_\mu = (\mathcal{A}_\mu^{(B)} + \mathcal{A}_\mu^{(K)}) - i\mathcal{A}_\mu^{(S)}$ is strictly gauge-invariant:
\begin{equation}\label{eq:app_amu_gauge}
    \bar{\mathcal{A}}_\mu = \mathcal{A}_\mu.
\end{equation}
It follows immediately that the complex non-Hermitian Berry phase $\gamma = \oint \mathcal{A}_\mu d\lambda^\mu$ is a gauge-invariant under the Dyson gauge freedom and therefore a well-defined physical observable.\\
Since, $\ket{\sigma_\mu}$ transforms covariantly and $\mathcal{A}_\mu^{(S)}$ is gauge-invariant, the stretching scalar $\mathcal{I}_\mu$ defined in Eq.~\eqref{eq:stretching_scalar} is gauge-invariant:
\begin{equation}\label{eq:I_gauge}
    \bar{\mathcal{I}}_\mu = \bigl(\bar{\mathcal{A}}_\mu^{(S)}\bigr)^2 + \braket{\bar{\sigma}_\mu}{\bar{\sigma}_\mu} = \bigl(\mathcal{A}_\mu^{(S)}\bigr)^2 + \braket{\sigma_\mu}{\sigma_\mu} = \mathcal{I}_\mu.
\end{equation}
Similarly, the system-level scalar $\operatorname{tr}(S_\mu S_\nu)$ is gauge-invariant due to the cyclic property of the trace:
\begin{equation}
    \operatorname{tr}(\bar{S}_\mu \bar{S}_\nu) = \operatorname{tr}(U S_\mu S_\nu U^\dagger) = \operatorname{tr}(S_\mu S_\nu).
\end{equation}
In contrast, the rotation scalar $\mathcal{J}_\mu$ [Eq.~\eqref{eq:j_scalar}] is constructed from gauge-dependent components ($\mathcal{A}_\mu^{(K)}$ and $\ket{\kappa_\mu}$). The resulting gauge transformations of the projected quantities are:
\begin{align}
    \bar{\mathcal{J}}_\mu &= \bigl(\mathcal{A}_\mu^{(K)} + a_\mu\bigr)^2 + \bigl(\bra{\kappa_\mu}U^\dagger + \bra{\tilde{\omega}_\mu}U^\dagger\bigr)\bigl(U\ket{\kappa_\mu} + U\ket{\tilde{\omega}_\mu}\bigr) \nonumber \\
    &= \mathcal{J}_\mu + 2 a_\mu \mathcal{A}_\mu^{(K)} + a_\mu^2 + 2\operatorname{Re}\braket{\kappa_\mu}{\tilde{\omega}_\mu} + \|\tilde{\omega}_\mu\|^2.
\end{align}
The presence of the gauge-shift scalar $a_\mu$ and the gauge fluctuation state $\ket{\tilde{\omega}_\mu}$ in the above equation explicitly demonstrates that $\mathcal{J}_\mu$ transforms inhomogeneously and is therefore a gauge-dependent quantity.
\subsection{Curvatures and the QGT}
%
The scalar curvatures [Eqs.~\eqref{eq:omega}--\eqref{eq:FS}] transform as:
\begin{align}
    \bar{\Omega}_{\mu\nu}^{(B)} &= \Omega_{\mu\nu}^{(B)} - f_{\mu\nu}, \\
    \bar{\Omega}_{\mu\nu}^{(K)} &= \Omega_{\mu\nu}^{(K)} + f_{\mu\nu}, \\
    \bar{\Omega}_{\mu\nu}^{(S)} &= \Omega_{\mu\nu}^{(S)},
\end{align}
where $f_{\mu\nu} \equiv \partial_\mu a_\nu - \partial_\nu a_\mu$. The standard Berry curvature and the rotation curvature are individually gauge-dependent, their sum, the total phase curvature $\Omega_{\mu\nu}^{(B)} + \Omega_{\mu\nu}^{(K)}$, is strictly gauge-invariant. Consequently, the full non-Hermitian Berry curvature $\Omega_{\mu\nu} = (\Omega_{\mu\nu}^{(B)} + \Omega_{\mu\nu}^{(K)}) - i \Omega_{\mu\nu}^{(S)}$ is gauge-invariant.\\
The non-Hermitian QGT components ($g_{\mu\nu}$ and $\Omega_{\mu\nu}$) are constructed exclusively from inner products of the covariant states $\ket{\Phi_\mu}$ and $\ket{\sigma_\mu}$ (see Eqs.~\eqref{eq:re_g}--\eqref{eq:im_omega}). Because the inner product of any two covariant states is invariant (e.g., $\braket{\bar{\Phi}_\mu}{\bar{\sigma}_\nu} = \bra{\Phi_\mu} U^\dagger U \ket{\sigma_\nu} = \braket{\Phi_\mu}{\sigma_\nu}$), every component of the non-Hermitian QGT is manifestly gauge-invariant:
\begin{equation}
    \bar{g}_{\mu\nu} = g_{\mu\nu}, \qquad \bar{\Omega}_{\mu\nu} = \Omega_{\mu\nu}.
\end{equation}
%
\subsection{Field Strengths}

The operator-valued field strengths defined in Sec.~\ref{sec:complex_berry_phase} transform covariantly as adjoint fields. The rotation field strength $\mathcal{F}_{\mu\nu}^{(K)}$ [Eq.~\eqref{eq:K_field_strength}], transforms as:
\begin{equation}
    \bar{\mathcal{F}}_{\mu\nu}^{(K)} = U \mathcal{F}_{\mu\nu}^{(K)} U^\dagger.
\end{equation}
For the stretching field strength $\mathcal{F}_{\mu\nu}^{(S)}$ [Eq.~\eqref{eq:S_field_strength}], the covariance follows directly from the transformation property of the adjoint $K$-covariant derivative given in Eq.~\eqref{eq:dd}. We obtain:
\begin{equation}
    \bar{\mathcal{F}}_{\mu\nu}^{(S)} = U \mathcal{F}_{\mu\nu}^{(S)} U^\dagger.
\end{equation}
The full Dyson connection curvature $\mathcal{F}_{\mu\nu}^{(\Gamma)}$ similarly transforms as:
\begin{equation}
    \bar{\mathcal{F}}_{\mu\nu}^{(\Gamma)} = \bar{\mathcal{F}}_{\mu\nu}^{(K)} + \bar{\mathcal{F}}_{\mu\nu}^{(S)} - [\bar{S}_\mu, \bar{S}_\nu] = U \mathcal{F}_{\mu\nu}^{(\Gamma)} U^\dagger.
\end{equation}
\subsection{Classification of quantities}
The geometric quantities fall into three transformation classes, summarized in Table~\ref{tab:gauge_classification}. Physical observables must be gauge-invariant, i.e., independent of the choice of the Dyson map.
\vspace{2em}
\begin{table}[htp]
    \centering
    \renewcommand{\arraystretch}{1.3}
    \begin{tabular}{ll}
        \hline\hline
        \textbf{Quantity} & \textbf{Symbol(s)} \\
        \hline
        \multicolumn{2}{l}{\textbf{Class 1: Gauge-Invariant}} \\
        \multicolumn{2}{l}{(Physical Observables: $\bar{O} = O$)} \\
        \hline
        Metric operator,  metric deformation operator & $G$, $C_\mu$ \\
        Left and right eigenstates of $\mathcal{H}$ & $\bra{L}$, $\ket{R}$ \\
        Stretching projection and curvature & $\mathcal{A}_\mu^{(S)}$, $F_{\mu\nu}^{(S)}$ \\
        Total phase connection & $\mathcal{A}_\mu^{(B)} + \mathcal{A}_\mu^{(K)}$ \\
        Total phase curvature & $\Omega_{\mu\nu}^{(B)} + F_{\mu\nu}^{(K)}$ \\
        Non-Hermitian Berry connection & $\mathcal{A}_\mu$\\
        Non-Hermitian Berry phase & $\gamma$ \\
        Stretching scalar and system-level scalar & $\mathcal{I}_\mu$, $\operatorname{tr}(S_\mu S_\nu)$ \\
        All QGT components & $g_{\mu\nu}$, $\Omega_{\mu\nu}$ \\
        \hline
        \multicolumn{2}{l}{\textbf{Class 2: Gauge-Covariant}} \\
        \multicolumn{2}{l}{(Transform as $X \to UXU^\dagger$ and $\ket{\psi} \to U\ket{\psi}$)} \\
        \hline
        Equivalent Hermitian Hamiltonian & $H$ \\
        Orthogonal projector & $\mathcal{Q}$ \\
        Hermitian eigenstates & $\ket{\Psi_H}$ \\
        Stretching component & $S_\mu$ \\
        Stretching fluctuation state & $\ket{\sigma_\mu}$ \\
        Dressed derivative state & $\ket{\Phi_\mu}$ \\
        Covariant derivatives & $D_\mu$, $D_\mu^{(K)}$,\\ & $\mathcal{D}_\mu^{(K)}$ \\
        Operator-valued field strengths & $\mathcal{F}_{\mu\nu}^{(K)}$, $\mathcal{F}_{\mu\nu}^{(S)}$,\\ & $\mathcal{F}_{\mu\nu}^{(\Gamma)}$ \\
        \hline
        \multicolumn{2}{l}{\textbf{Class 3: Gauge-Dependent}} \\
        \multicolumn{2}{l}{(Acquire inhomogeneous pure-gauge artifacts)} \\
        \hline
        Dyson map & $\eta$ \\
        Dyson connection & $\Gamma_\mu$ \\
        Rotation component & $K_\mu$ \\
        State derivative & $\ket{\phi_\mu}$ \\
        Rotation fluctuation state & $\ket{\kappa_\mu}$ \\
        Standard Berry connection and curvature & $\mathcal{A}_\mu^{(B)}$, $\Omega_{\mu\nu}^{(B)}$ \\
        Rotation projection and curvature & $\mathcal{A}_\mu^{(K)}$, $F_{\mu\nu}^{(K)}$ \\
        Rotation scalar & $\mathcal{J}_\mu$ \\
        \hline\hline
    \end{tabular}
    \caption{Classification of geometric quantities under the Dyson gauge freedom $\eta \to \bar{\eta}$.}
    \label{tab:gauge_classification}
\end{table}
\section{Connecting Abelian Curls to covariant states}
\label{app:identities}

In this appendix, we derive the identities presented in Eqs.~\eqref{eq:identity_Re} and~\eqref{eq:identity_Im}, which relate the Abelian field strengths defined via the curls of the scalar connections [Eqs.~\eqref{eq:omega}--\eqref{eq:FS}] to the inner products of the covariant projected states $\ket{\Phi_\mu}$ and $\ket{\sigma_\mu}$.
%
\subsection{Projected Inner Products}
%
We begin by evaluating the inner products of the orthogonal projected states defined in Sec.~\ref{sec:decomposition}: $\ket{\phi_\mu} = \mathcal{Q}\ket{\partial_\mu\Psi_H}$, $\ket{\kappa_\mu} = \mathcal{Q}K_\mu\ket{\Psi_H}$, and $\ket{\sigma_\mu} = \mathcal{Q}S_\mu\ket{\Psi_H}$. Inserting the projector $\mathcal{Q} = \mathbb{I} - \ket{\Psi_H}\bra{\Psi_H}$ and utilizing the Hermiticity properties $S_\mu^\dagger = S_\mu$ and $K_\mu^\dagger = -K_\mu$, the nine fundamental inner products comprising the QGT are:
\begin{subequations}\label{eq:app_nine_inner}
\begin{align}
  \braket{\phi_\mu}{\phi_\nu}
    &= \braket{\partial_\mu\Psi_H}{\partial_\nu\Psi_H} - \mathcal{A}^{(B)}_\mu\mathcal{A}^{(B)}_\nu, \\[4pt]
  \braket{\phi_\mu}{\kappa_\nu}
    &= \mel{\partial_\mu\Psi_H}{K_\nu}{\Psi_H} + \mathcal{A}^{(B)}_\mu\mathcal{A}^{(K)}_\nu, \\[4pt]
  \braket{\phi_\mu}{\sigma_\nu}
    &= \mel{\partial_\mu\Psi_H}{S_\nu}{\Psi_H} - i\,\mathcal{A}^{(B)}_\mu\mathcal{A}^{(S)}_\nu, \\[4pt]
  \braket{\kappa_\mu}{\phi_\nu}
    &= -\mel{\Psi_H}{K_\mu}{\partial_\nu\Psi_H} + \mathcal{A}^{(K)}_\mu\mathcal{A}^{(B)}_\nu, \\[4pt]
  \braket{\sigma_\mu}{\phi_\nu}
    &= \mel{\Psi_H}{S_\mu}{\partial_\nu\Psi_H} + i\,\mathcal{A}^{(S)}_\mu\mathcal{A}^{(B)}_\nu, \\[4pt]
  \braket{\kappa_\mu}{\kappa_\nu}
    &= -\mel{\Psi_H}{K_\mu K_\nu}{\Psi_H} - \mathcal{A}^{(K)}_\mu\mathcal{A}^{(K)}_\nu, \\[4pt]
  \braket{\sigma_\mu}{\sigma_\nu}
    &= \mel{\Psi_H}{S_\mu S_\nu}{\Psi_H} - \mathcal{A}^{(S)}_\mu\mathcal{A}^{(S)}_\nu, \\[4pt]
  \braket{\kappa_\mu}{\sigma_\nu}
    &= -\mel{\Psi_H}{K_\mu S_\nu}{\Psi_H} + i\,\mathcal{A}^{(K)}_\mu\mathcal{A}^{(S)}_\nu, \\[4pt]
  \braket{\sigma_\mu}{\kappa_\nu}
    &= \mel{\Psi_H}{S_\mu K_\nu}{\Psi_H} - i\,\mathcal{A}^{(S)}_\mu\mathcal{A}^{(K)}_\nu.
\end{align}
\end{subequations}
\subsection{Real part of Curvature }
%
To establish Eq.~\eqref{eq:identity_Re}, we start with the right-hand side by substituting $\ket{\Phi_\mu} = \ket{\phi_\mu} - \ket{\kappa_\mu}$. The right-hand side of the identity requires the imaginary parts of the inner products. Since the scalar connections $\mathcal{A}_\mu^{(B)}$, $\mathcal{A}_\mu^{(K)}$, and $\mathcal{A}_\mu^{(S)}$ are strictly real, their products do not contribute to the imaginary parts. Extracting the imaginary components from Eq.~\eqref{eq:app_nine_inner} and applying $K_\mu^\dagger = -K_\mu$ to simplify cross-terms (e.g., $\mathrm{Im}\mel{\Psi_H}{K_\mu}{\partial_\nu\Psi_H} = \mathrm{Im}\mel{\partial_\nu\Psi_H}{K_\mu}{\Psi_H}$), we obtain:
\begin{equation}\label{eq:app_RHS_Re}
\begin{aligned}
  &-2\,\mathrm{Im}\braket{\Phi_\mu}{\Phi_\nu} + 2\,\mathrm{Im}\braket{\sigma_\mu}{\sigma_\nu} \\
  &\quad= -2\,\mathrm{Im}\braket{\partial_\mu\Psi_H}{\partial_\nu\Psi_H} 
       + 2\,\mathrm{Im}\mel{\partial_\mu\Psi_H}{K_\nu}{\Psi_H} \\
  &\qquad - 2\,\mathrm{Im}\mel{\partial_\nu\Psi_H}{K_\mu}{\Psi_H} 
       + 2\,\mathrm{Im}\mel{\Psi_H}{K_\mu K_\nu}{\Psi_H} \\
  &\qquad + 2\,\mathrm{Im}\mel{\Psi_H}{S_\mu S_\nu}{\Psi_H}.
\end{aligned}
\end{equation}
The first term is precisely the standard Berry curvature $\Omega_{\mu\nu}^{(B)}$ defined in Eq.~\eqref{eq:omega}. We now evaluate $\Omega_{\mu\nu}^{(K)} = \partial_\mu\mathcal{A}_\nu^{(K)} - \partial_\nu\mathcal{A}_\mu^{(K)}$ in the left-hand side of Eq.~\eqref{eq:identity_Re}, directly from the definition $\mathcal{A}_\mu^{(K)} = -i\mel{\Psi_H}{K_\mu}{\Psi_H}$. Expanding the parameter derivatives and antisymmetrizing yields:
\begin{equation}\label{eq:app_FK_expansion}
\begin{aligned}
  \Omega_{\mu\nu}^{(K)} 
  &= 2\,\mathrm{Im}\mel{\partial_\mu\Psi_H}{K_\nu}{\Psi_H} 
   - 2\,\mathrm{Im}\mel{\partial_\nu\Psi_H}{K_\mu}{\Psi_H} \\
  &\quad -i\mel{\Psi_H}{(\partial_\mu K_\nu - \partial_\nu K_\mu)}{\Psi_H}.
\end{aligned}
\end{equation}
The derivative of the operators is resolved by the zero-curvature condition of the Dyson connection, $\mathcal{F}_{\mu\nu}^{(\Gamma)} = 0$ [Eq.~\eqref{eq:flatness}]. The anti-Hermitian part of zero curvature condition provides the operator identity:
\begin{equation}\label{eq:app_K_curl}
  \partial_\mu K_\nu - \partial_\nu K_\mu = [S_\mu, S_\nu] + [K_\mu, K_\nu].
\end{equation}
Noting that the expectation value of the commutator of any two Hermitian (or anti-Hermitian) operators evaluates to $2i$ times the imaginary part of their product, substituting Eq.~\eqref{eq:app_K_curl} into Eq.~\eqref{eq:app_FK_expansion} gives:
\begin{equation}\label{eq:app_FK_final}
\begin{aligned}
  \Omega_{\mu\nu}^{(K)} 
  &= 2\,\mathrm{Im}\mel{\partial_\mu\Psi_H}{K_\nu}{\Psi_H} 
   - 2\,\mathrm{Im}\mel{\partial_\nu\Psi_H}{K_\mu}{\Psi_H} \\
  &\quad + 2\,\mathrm{Im}\mel{\Psi_H}{S_\mu S_\nu}{\Psi_H} 
   + 2\,\mathrm{Im}\mel{\Psi_H}{K_\mu K_\nu}{\Psi_H}.
\end{aligned}
\end{equation}
Substituting this in the left-hand side of Eq~\eqref{eq:identity_Re}, we reproduce Eq.~\eqref{eq:app_RHS_Re}, thereby proving the identity given in Eq.~\eqref{eq:identity_Re}
%
\subsection{Imaginary part of Curvature }
%
To establish Eq.~\eqref{eq:identity_Im}, we follow an analogous procedure for the real parts of the cross-terms. Expanding the right-hand side, $2\,\mathrm{Re}\braket{\Phi_\mu}{\sigma_\nu} - 2\,\mathrm{Re}\braket{\sigma_\mu}{\Phi_\nu}$, and utilizing Eq.~\eqref{eq:app_nine_inner}. Since terms containing a single factor of $i$ alongside real connections are purely imaginary, they drop out of the real part. Applying operator Hermiticity to simplify the remaining terms yields:
\begin{equation}\label{eq:app_RHS_Im}
\begin{aligned}
  &2\,\mathrm{Re}\braket{\Phi_\mu}{\sigma_\nu} - 2\,\mathrm{Re}\braket{\sigma_\mu}{\Phi_\nu} \\
  &\quad= 2\,\mathrm{Re}\mel{\partial_\mu\Psi_H}{S_\nu}{\Psi_H} 
       - 2\,\mathrm{Re}\mel{\partial_\nu\Psi_H}{S_\mu}{\Psi_H} \\
  &\qquad + 2\,\mathrm{Re}\mel{\Psi_H}{K_\mu S_\nu}{\Psi_H} 
       - 2\,\mathrm{Re}\mel{\Psi_H}{K_\nu S_\mu}{\Psi_H}.
\end{aligned}
\end{equation}

For the left-hand side, we expand $\Omega_{\mu\nu}^{(S)} = \partial_\mu\mathcal{A}_\nu^{(S)} - \partial_\nu\mathcal{A}_\mu^{(S)}$ using $\mathcal{A}_\mu^{(S)} = \mel{\Psi_H}{S_\mu}{\Psi_H}$:
\begin{equation}\label{eq:app_FS_expansion}
\begin{aligned}
  \Omega_{\mu\nu}^{(S)} 
  &= 2\,\mathrm{Re}\mel{\partial_\mu\Psi_H}{S_\nu}{\Psi_H} 
   - 2\,\mathrm{Re}\mel{\partial_\nu\Psi_H}{S_\mu}{\Psi_H} \\
  &\quad + \mel{\Psi_H}{(\partial_\mu S_\nu - \partial_\nu S_\mu)}{\Psi_H}.
\end{aligned}
\end{equation}
The Hermitian part of the zero-curvature condition yields the companion operator identity:
\begin{equation}\label{eq:app_S_curl}
  \partial_\mu S_\nu - \partial_\nu S_\mu = [S_\mu, K_\nu] + [K_\mu, S_\nu].
\end{equation}
Evaluating the expectation value of these mixed commutators requires care. Since $(S_\mu K_\nu)^\dagger = -K_\nu S_\mu$, it follows that $\mel{\Psi_H}{[S_\mu, K_\nu]}{\Psi_H} = 2\,\mathrm{Re}\mel{\Psi_H}{S_\mu K_\nu}{\Psi_H} = -2\,\mathrm{Re}\mel{\Psi_H}{K_\nu S_\mu}{\Psi_H}$. Substituting Eq.~\eqref{eq:app_S_curl} into Eq.~\eqref{eq:app_FS_expansion} and applying this trace property gives:
\begin{equation}
\begin{aligned}
  \Omega_{\mu\nu}^{(S)} 
  &= 2\,\mathrm{Re}\mel{\partial_\mu\Psi_H}{S_\nu}{\Psi_H} 
   - 2\,\mathrm{Re}\mel{\partial_\nu\Psi_H}{S_\mu}{\Psi_H} \\
  &\quad - 2\,\mathrm{Re}\mel{\Psi_H}{K_\nu S_\mu}{\Psi_H} 
   + 2\,\mathrm{Re}\mel{\Psi_H}{K_\mu S_\nu}{\Psi_H}.
\end{aligned}
\end{equation}
This expression is identical to Eq.~\eqref{eq:app_RHS_Im}, which proves the identity presented in Eq.~\eqref{eq:identity_Im}.
%
\section{Off-diagonal matrix elements of \texorpdfstring{$S_\mu$}{Smu},
\texorpdfstring{$K_\mu$}{Kmu}, and the covariant states}
\label{app:Smu_offdiag}
%
This appendix derives the off-diagonal matrix elements of the stretching component $S_\mu$ and the rotation component $K_\mu$ in the eigenbasis of the Hermitian Hamiltonian $H$. These results provide the spectral representations of the covariant states $\ket{\sigma_\mu}$ and $\ket{\Phi_\mu}$ used in Sec.~\ref{sec:ep_singularity}.

\subsection{Off-diagonal matrix elements of \texorpdfstring{$S_\mu$}{Smu} and \texorpdfstring{$K_\mu$}{Kmu}}

The parametric evolution of the equivalent Hermitian Hamiltonian $H = \eta \mathcal{H} \eta^{-1}$ is governed by its derivative with respect to $\lambda^\mu$:
\begin{equation}\label{eq:app_dh}
    \partial_\mu H = [\Gamma_\mu, H] + \eta (\partial_\mu \mathcal{H}) \eta^{-1}\,.
\end{equation}
We evaluate the matrix elements of both sides between distinct eigenstates $\bra{\Psi_H^{(n)}}$ and $\ket{\Psi_H^{(m)}}$ of $H$ ($n \neq m$). Expanding $H = \sum_p E_p \ket{\Psi_H^{(p)}}\bra{\Psi_H^{(p)}}$ and applying the orthonormality condition $\braket{\Psi_H^{(n)}}{\Psi_H^{(m)}} = \delta_{nm}$, the derivative evaluates to:
\begin{equation}\label{eq:app_lhs_final}
    \mel{\Psi_H^{(n)}}{\partial_\mu H}{\Psi_H^{(m)}} = (E_m - E_n) \braket{\Psi_H^{(n)}}{\partial_\mu \Psi_H^{(m)}}.
\end{equation}
Similarly, the commutator term gives:
\begin{equation}
    \mel{\Psi_H^{(n)}}{[\Gamma_\mu, H]}{\Psi_H^{(m)}} = (E_m - E_n) \mel{\Psi_H^{(n)}}{\Gamma_\mu}{\Psi_H^{(m)}}.
\end{equation}
Substituting these results into the Eq.~\eqref{eq:app_dh} and dividing by the non-zero energy gap $E_m - E_n$, we obtain the off-diagonal elements of the Dyson connection:
\begin{align}\label{eq:app_Gamma_offdiag}
    \mel{\Psi_H^{(n)}}{\Gamma_\mu}{\Psi_H^{(m)}} 
    &= \braket{\Psi_H^{(n)}}{\partial_\mu \Psi_H^{(m)}} \notag \\
    &\quad - \frac{\mel{\Psi_H^{(n)}}{\eta (\partial_\mu \mathcal{H}) \eta^{-1}}{\Psi_H^{(m)}}}{E_m - E_n}.
\end{align}
Following an analogous procedure for the Hermitian conjugate $H = H^\dagger$, the off-diagonal elements of $\Gamma_\mu^\dagger$ are:
\begin{align}\label{eq:app_Gammadagger_offdiag}
    \mel{\Psi_H^{(n)}}{\Gamma_\mu^\dagger}{\Psi_H^{(m)}} 
    &= -\braket{\Psi_H^{(n)}}{\partial_\mu \Psi_H^{(m)}} \notag \\
    &\quad + \frac{\mel{\Psi_H^{(n)}}{\eta^{-\dagger} (\partial_\mu \mathcal{H}^\dagger) \eta^\dagger}{\Psi_H^{(m)}}}{E_m - E_n}.
\end{align}
Adding Eqs.~\eqref{eq:app_Gamma_offdiag} and \eqref{eq:app_Gammadagger_offdiag}, the terms $\braket{\Psi_H^{(n)}}{\partial_\mu \Psi_H^{(m)}}$ cancel exactly, and we obtain the off-diagonal matrix elements of $S_\mu$ given in Eq.~\eqref{eq:Smu_offdiag}. Similarly, subtracting Eq.~\eqref{eq:app_Gammadagger_offdiag} from Eq.~\eqref{eq:app_Gamma_offdiag}, the terms $\braket{\Psi_H^{(n)}}{\partial_\mu \Psi_H^{(m)}}$ survive. The off-diagonal matrix element of $K_\mu$ becomes:
\begin{equation}\label{eq:app_Kmu_compact}
    \mel{\Psi_H^{(n)}}{K_\mu}{\Psi_H^{(m)}} = \braket{\Psi_H^{(n)}}{\partial_\mu \Psi_H^{(m)}} - \frac{\mathcal{N}_\mu^{(nm)}}{E_m - E_n}, \qquad n \neq m,
\end{equation}
with $\mathcal{N}_\mu^{(nm)}$ defined in Eq.~\eqref{eq:N_amplitude}.

\subsection{Spectral representation of \texorpdfstring{$\ket{\sigma_\mu}$}{sigma\_mu}}

The stretching fluctuation state $\ket{\sigma_\mu}$ [Eq.~\eqref{eq:sigma}], can be expanded by inserting the completeness relation $\sum_n \ket{\Psi_H^{(n)}}\bra{\Psi_H^{(n)}} = \mathbb{I}$, yielding the spectral representation:
\begin{equation}\label{eq:app_sigma_expand}
    \ket{\sigma_\mu} = \sum_{n \neq m} \frac{\mathcal{M}_\mu^{(nm)}}{E_m - E_n} \ket{\Psi_H^{(n)}}.
\end{equation}
The squared norm and the inner product between stretching fluctuations state along two parameter directions are given respectively by:
\begin{align}
    \|\sigma_\mu\|^2 &= \sum_{n \neq m} \frac{|\mathcal{M}_\mu^{(nm)}|^2}{|E_m - E_n|^2}, \label{eq:app_sigma_norm} \\
    \braket{\sigma_\mu}{\sigma_\nu} &= \sum_{n \neq m} \frac{\mathcal{M}_\mu^{(nm)*} \mathcal{M}_\nu^{(nm)}}{|E_m - E_n|^2}. \label{eq:app_sigma_inner}
\end{align}
\subsection{Spectral representation of \texorpdfstring{$\ket{\Phi_\mu}$}{Phi\_mu}}
Projecting the dressed derivative state $\ket{\Phi_\mu}$ [Eq.~\eqref{eq:Phi}] onto an orthogonal eigenstate $\bra{\Psi_H^{(n)}}$ ($n \neq m$) yields its matrix elements in the eigenbasis of $H$:
\begin{equation}
    \mel{\Psi_H^{(n)}}{D_\mu^{(K)}}{\Psi_H^{(m)}} = \braket{\Psi_H^{(n)}}{\partial_\mu \Psi_H^{(m)}} - \mel{\Psi_H^{(n)}}{K_\mu}{\Psi_H^{(m)}}.
\end{equation}
Substituting the off-diagonal matrix element of $K_\mu$ from Eq.~\eqref{eq:app_Kmu_compact}, the terms $\braket{\Psi_H^{(n)}}{\partial_\mu \Psi_H^{(m)}}$ cancel exactly, recovering the matrix elements of $D_\mu^{(K)}$ given in Eq.~\eqref{eq:Phi_matrix_element}. Inserting the completeness relation provides the spectral expansion for the dressed derivative state:
\begin{equation}\label{eq:app_Phi_expand}
    \ket{\Phi_\mu} = \sum_{n \neq m} \frac{\mathcal{N}_\mu^{(nm)}}{E_m - E_n} \ket{\Psi_H^{(n)}}.
\end{equation}
The corresponding squared norm and inner product take the form:
\begin{align}
    \|\Phi_\mu\|^2 &= \sum_{n \neq m} \frac{|\mathcal{N}_\mu^{(nm)}|^2}{|E_m - E_n|^2}, \label{eq:app_Phi_norm} \\
    \braket{\Phi_\mu}{\Phi_\nu} &= \sum_{n \neq m} \frac{\mathcal{N}_\mu^{(nm)*} \mathcal{N}_\nu^{(nm)}}{|E_m - E_n|^2}. \label{eq:app_Phi_inner}
\end{align}
\subsection{Cross inner product}
The cross inner product between the dressed derivative and the stretching fluctuation states, which dictates the imaginary parts of the non-Hermitian QGT, follows directly from Eqs.~\eqref{eq:app_sigma_expand} and \eqref{eq:app_Phi_expand}:
\begin{equation}\label{eq:app_cross_inner}
    \braket{\Phi_\mu}{\sigma_\nu} = \sum_{n \neq m} \frac{\mathcal{N}_\mu^{(nm)*} \mathcal{M}_\nu^{(nm)}}{|E_m - E_n|^2}.
\end{equation}
All three covariant inner products ($\braket{\Phi_\mu}{\Phi_\nu}$, $\braket{\sigma_\mu}{\sigma_\nu}$, and $\braket{\Phi_\mu}{\sigma_\nu}$) share the same spectral structure; a sum over off-diagonal eigenstates weighted by $|E_m - E_n|^{-2}$.
%
%
\section{The RR and LL Quantum Geometric Tensors and Their Gauge Structure}
\label{app:RRLL}
%
%
%
In Sec.~\ref{sec:qgt}, the non-Hermitian QGT was constructed in the biorthogonal (LR) definition. Here we derive the right--right (RR) and left--left (LL) forms in the same covariant basis to examine which form is most naturally suited to the gauge-covariant framework developed in this work.
%
\subsection{Reduction to the covariant basis}
\label{app:frame_deriv}
%
Differentiating $\ket{R}=\eta^{-1}\ket{\Psi_H}$ and $\ket{L}=\eta^{\dagger}\ket{\Psi_H}$ with respect to $\lambda^\mu$ yields
\begin{equation}\label{eq:dRL}
	\ket{\partial_\mu R}=\eta^{-1}\ket{V_\mu^{R}},
	\qquad
	\ket{\partial_\mu L}=\eta^{\dagger}\ket{V_\mu^{L}},
\end{equation}
where
\begin{equation}\label{eq:V_def}
	\ket{V_\mu^{R}}\equiv(\partial_\mu-\Gamma_\mu)\ket{\Psi_H},
	\qquad
	\ket{V_\mu^{L}}\equiv(\partial_\mu+\Gamma_\mu^{\dagger})\ket{\Psi_H}.
\end{equation}
Their orthogonal projections onto the covariant states are
\begin{subequations}\label{eq:chixi}
	\begin{align}
		\mathcal{Q}\ket{V_\mu^{R}} &= \ket{\Phi_\mu}-\ket{\sigma_\mu}
		\equiv\ket{\chi_\mu}, \label{eq:chi_def}\\
		\mathcal{Q}\ket{V_\mu^{L}} &= \ket{\Phi_\mu}+\ket{\sigma_\mu}
		\equiv\ket{\xi_\mu}. \label{eq:xi_def}
	\end{align}
\end{subequations}
the sign flip on $\ket{\sigma_\mu}$ arising from $\Gamma_\mu^{\dagger}=S_\mu-K_\mu$. The full decompositions are
\begin{equation}\label{eq:V_decomp}
	\ket{V_\mu^{R}}=r_\mu\ket{\Psi_H}+\ket{\chi_\mu},
	\qquad
	\ket{V_\mu^{L}}=\ell_\mu\ket{\Psi_H}+\ket{\xi_\mu},
\end{equation}
with the parallel projections
\begin{align}
	r_\mu &\equiv \mel{\Psi_H}{(\partial_\mu-\Gamma_\mu)}{\Psi_H}
	= -i\mathcal{A}_\mu^{(B)}-\mathcal{A}_\mu^{(S)}-i\mathcal{A}_\mu^{(K)},
	\label{eq:r_def}\\
	\ell_\mu &\equiv
	\mel{\Psi_H}{(\partial_\mu+\Gamma_\mu^{\dagger})}{\Psi_H}
	= -i\mathcal{A}_\mu^{(B)}+\mathcal{A}_\mu^{(S)}-i\mathcal{A}_\mu^{(K)}.
	\label{eq:l_def}
\end{align}
%
\subsection{The RR-QGT}
\label{app:RR}
%
The RR definition uses only the right states with RR projector $\mathbb{I}-\ket{R}\bra{R}/\braket{R}{R}$,
\begin{equation}\label{eq:QRR_def}
	Q^{RR}_{\mu\nu}
	=\braket{\partial_\mu R}{\partial_\nu R}
	-\frac{\braket{\partial_\mu R}{R}\,\braket{R}{\partial_\nu R}}{\braket{R}{R}}.
\end{equation}
The right state is not normalized in the Dirac sense, gives
\begin{equation}\label{eq:nR}
	\braket{R}{R}
	=\mel{\Psi_H}{\eta^{-\dagger}\eta^{-1}}{\Psi_H}
	=\mel{\Psi_H}{\widetilde{G}^{-1}}{\Psi_H}
\end{equation}
where $\widetilde{G}\equiv \eta\eta^\dagger$. Using Eq.~\eqref{eq:dRL}, we can write,
\begin{subequations}\label{eq:RR_overlaps}
	\begin{align}
		\braket{\partial_\mu R}{\partial_\nu R}
		&=\mel{V_\mu^{R}}{\widetilde{G}^{-1}}{V_\nu^{R}},
		\label{eq:RR_overlap_dd}\\
		\braket{\partial_\mu R}{R}
		&=\mel{V_\mu^{R}}{\widetilde{G}^{-1}}{\Psi_H}.
		\label{eq:RR_overlap_d}
	\end{align}
\end{subequations}
Substituting Eq.~\eqref{eq:V_decomp} into Eq.~\eqref{eq:RR_overlaps} and inserting the result into Eq.~\eqref{eq:QRR_def}, the RR definition reduces to
\begin{equation}\label{eq:QRR_final}
	Q^{RR}_{\mu\nu}
	=\mel{\chi_\mu}{\widetilde{G}^{-1}}{\chi_\nu}
	-\frac{\mel{\chi_\mu}{\widetilde{G}^{-1}}{\Psi_H}\,
		\mel{\Psi_H}{\widetilde{G}^{-1}}{\chi_\nu}}
	{\mel{\Psi_H}{\widetilde{G}^{-1}}{\Psi_H}}.
\end{equation}
%
\subsection{The LL and LR-QGT}
\label{app:LL}
%
Repeating the same steps with $\ket{R}\to \ket{L}$, i.e., $\ket{V_\mu^{R}}\to\ket{V_\mu^{L}}$ and $\widetilde{G}^{-1}\to\widetilde{G}$, gives
\begin{equation}\label{eq:QLL_final}
	Q^{LL}_{\mu\nu}
	=\mel{\xi_\mu}{\widetilde{G}}{\xi_\nu}
	-\frac{\mel{\xi_\mu}{\widetilde{G}}{\Psi_H}\,
		\mel{\Psi_H}{\widetilde{G}}{\xi_\nu}}
	{\mel{\Psi_H}{\widetilde{G}}{\Psi_H}},
\end{equation}
For reference, the LR form given in Eq.~\eqref{eq:QGT_biorthogonal} is recovered as
\begin{equation}\label{eq:LR_check}
	Q_{\mu\nu}
	=\bra{V_\mu^{L}}\mathcal{Q}\ket{V_\nu^{R}}
	=\braket{\xi_\mu}{\chi_\nu},
\end{equation}
The three forms [Eqs.~\eqref{eq:QRR_final},~\eqref{eq:QLL_final}, and~\eqref{eq:LR_check}] are summarized in Table~\ref{tab:weights}. In the polar gauge $\eta=G^{1/2}$,
$\widetilde{G}=G$, so the RR and LL-QGTs reduce to
\begin{align}
	Q^{RR}_{\mu\nu}
	&=\mel{\chi_\mu}{G^{-1}}{\chi_\nu}
	-\frac{\mel{\chi_\mu}{G^{-1}}{\Psi_H}\,
		\mel{\Psi_H}{G^{-1}}{\chi_\nu}}
	{\mel{\Psi_H}{G^{-1}}{\Psi_H}},
	\label{eq:QRR_polar}\\
	Q^{LL}_{\mu\nu}
	&=\mel{\xi_\mu}{G}{\xi_\nu}
	-\frac{\mel{\xi_\mu}{G}{\Psi_H}\,
		\mel{\Psi_H}{G}{\xi_\nu}}
	{\mel{\Psi_H}{G}{\Psi_H}}.
	\label{eq:QLL_polar}
\end{align}
Comparing the three forms, the RR and LL tensors carry an explicit weighting by $\widetilde{G}^{-1}$ and $\widetilde{G}$ respectively, measuring the geometry in the deformed inner product, whereas LR carries no such weighting (its Dyson-map factors cancel identically as $\eta^{\dagger}\left(\eta^{\dagger}\right)^{-1}=\mathbb{I}$ and $\eta\,\eta^{-1}=\mathbb{I}$), and the QGT is expressed purely through covariant states $\ket{\Phi_\mu}$ and $\ket{\sigma_\mu}$ in the standard inner product.
\begin{table}[h!]
	\centering
	\renewcommand{\arraystretch}{1.3}
	\begin{tabular*}{\columnwidth}{@{\extracolsep{\fill}}lcc}
		\hline\hline
		Form & Weight & Covariant State \\
		\hline
		LR & $\mathbb{I}$ & $\ket{\chi_\mu},\,\ket{\xi_\mu}$ \\
		RR & $\widetilde{G}^{-1}=(\eta\eta^{\dagger})^{-1}$
		& $\ket{\chi_\mu}$ \\
		LL & $\widetilde{G}=\eta\eta^{\dagger}$
		& $\ket{\xi_\mu}$ \\
		\hline\hline
	\end{tabular*}
	\caption{The three QGT forms in the covariant basis. Only LR form ($Q_{\mu\nu}$) carries no metric weighting.}
	\label{tab:weights}
\end{table}
%
\subsection{Gauge structure}
\label{app:gauge}
%
Under the Dyson gauge transformation $\eta\to \bar{\eta}$ the states $\ket{V_\mu^{R}}$ and $\ket{V_\mu^{L}}$ transformed covariantly as,
\begin{equation}\label{eq:Vbar}
	\begin{split}
		\ket{\bar V_\mu^{R}}
		&=(\partial_\mu-\bar\Gamma_\mu)\,U\ket{\Psi_H}\\
		&=\omega_\mu U\ket{\Psi_H}+U\ket{\partial_\mu\Psi_H}
		-\omega_\mu U\ket{\Psi_H}-U\Gamma_\mu\ket{\Psi_H}\\
		&=U(\partial_\mu-\Gamma_\mu)\ket{\Psi_H}=U\ket{V_\mu^{R}},
	\end{split}
\end{equation}
and, with $\bar\Gamma_\mu^{\dagger}=-\omega_\mu+U\Gamma_\mu^{\dagger}U^{\dagger}$, similarly
$\ket{\bar V_\mu^{L}}=U\ket{V_\mu^{L}}$. The operators $\widetilde{G}$ and $\widetilde{G}^{-1}$ transform covariantly as well,
\begin{equation}\label{eq:Gd_transf}
	\bar{\widetilde{G}}=\bar\eta\,\bar\eta^{\dagger}=U\widetilde{G}U^{\dagger},
	\qquad
	\bar{\widetilde{G}}^{-1}=U\widetilde{G}^{-1}U^{\dagger}.
\end{equation}
So every matrix element of $Q^{RR}_{\mu\nu}$ [Eq.~\eqref{eq:QRR_final}] satisfies
\begin{equation}\label{eq:RR_inv}
	\mel{\bar V_\mu^{R}}{\bar{\widetilde{G}}^{-1}}
	{\bar V_\nu^{R}}
	=\mel{V_\mu^{R}}{U^{\dagger}U\,\widetilde{G}^{-1}
		\,U^{\dagger}U}{V_\nu^{R}}
	=\mel{V_\mu^{R}}{\widetilde{G}^{-1}}{V_\nu^{R}},
\end{equation}
The same argument applies to $Q^{LL}_{\mu\nu}$ [Eq.~\eqref{eq:QLL_final}], giving
\begin{equation}\label{eq:QRRLL_inv}
	\bar Q^{RR}_{\mu\nu}=Q^{RR}_{\mu\nu},
	\qquad
	\bar Q^{LL}_{\mu\nu}=Q^{LL}_{\mu\nu}.
\end{equation}
The gauge invariance of all three forms confirms that the distinction noted in Table~\ref{tab:weights} is structural rather than gauge-theoretic. The absence of metric weighting in the LR form motivates its adoption as the natural non-Hermitian QGT in the Sec.~\ref{sec:qgt}.
\end{document}